\documentclass[acmsmall]{acmart}

\AtBeginDocument{%
	}

\acmJournal{JACM}
\acmVolume{xx}
\acmNumber{xx}
\acmArticle{xx}
\acmMonth{11}

\usepackage{amssymb}
\usepackage{amsmath}
\usepackage{amsfonts}
\usepackage{multirow}
\usepackage{graphicx}
\usepackage{textcomp}
\usepackage{amsthm}
\renewcommand{\vec}[1]{\boldsymbol{#1}}
\usepackage{setspace}
\usepackage{subfigure}
\usepackage{algorithmic}
\usepackage[ruled,vlined,boxed,linesnumbered]{algorithm2e}
\usepackage{color}

\newtheorem{defn}{Definition}

\begin{document}

\title{A Survey on Influence Maximization: From an ML-Based Combinatorial Optimization}

\author{Yandi Li}
\email{liyandi@uic.edu.cn}
\author{Haobo Gao}
\email{p930026030@mail.uic.edu.cn}
\author{Yunxuan Gao}
\email{p930026032@mail.uic.edu.cn}
\affiliation{%
  \institution{Department of Computer Science, BNU-HKBU United International College}
  \city{Zhuhai}
  \state{Guangdong}
  \country{China}
  \postcode{519087}
}

\author{Jianxiong Guo}
\email{jianxiongguo@bnu.edu.cn}
\authornote{Corresponding author}
\affiliation{%
  \institution{Advanced Institute of Natural Sciences, Beijing Normal University}
  \city{Zhuhai}
  \state{Guangdong}
  \country{China}
  \postcode{519087}
}
\affiliation{%
  \institution{Guangdong Key Lab of AI and Multi-Modal Data Processing, BNU-HKBU United International College}
  \city{Zhuhai}
  \state{Guangdong}
  \country{China}
  \postcode{519087}
}

\author{Weili Wu}
\email{weiliwu@utdallas.edu}
\affiliation{%
	\institution{Department of Computer Science, The University of Texas at Dallas}
	\streetaddress{800 W Campbell Rd}
	\city{Richardson}
	\state{Texas}
	\country{USA}
	\postcode{75080}
}

\renewcommand{\shortauthors}{Y. Li et al.}

\begin{abstract}
	Influence Maximization (IM) is a classical combinatorial optimization problem, which can be widely used in mobile networks, social computing, and recommendation systems. It aims at selecting a small number of users such that maximizing the influence spread across the online social network. Because of its potential commercial and academic value, there are a lot of researchers focusing on studying the IM problem from different perspectives. The main challenge comes from the NP-hardness of the IM problem and \#P-hardness of estimating the influence spread, thus traditional algorithms for overcoming them can be categorized into two classes: heuristic algorithms and approximation algorithms. However, there is no theoretical guarantee for heuristic algorithms, and the theoretical design is close to the limit. Therefore, it is almost impossible to further optimize and improve their performance.
	
	With the rapid development of artificial intelligence, the technology based on Machine Learning (ML) has achieved remarkable achievements in many fields. In view of this, in recent years, a number of new methods have emerged to solve combinatorial optimization problems by using ML-based techniques. These methods have the advantages of fast solving speed and strong generalization ability to unknown graphs, which provide a brand-new direction for solving combinatorial optimization problems. Therefore, we abandon the traditional algorithms based on iterative search and review the recent development of ML-based methods, especially Deep Reinforcement Learning, to solve the IM problem and other variants in social networks. We focus on summarizing the relevant background knowledge, basic principles, common methods, and applied research. Finally, the challenges that need to be solved urgently in future IM research are pointed out.
\end{abstract}



\maketitle

\section{Introduction}
With the rapid development of the mobile internet, social media platforms and content sharing platforms are rapidly emerging. Compared with newspapers, radio, television, and other traditional media, mobile Internet-based media has faster speed, lower cost, and more diverse forms. Thus, the commercial social giants represented by Weibo, Facebook, TikTok, and Twitter were quickly born and strengthened in the last two decades. All these applications are based on the Online Social Network (OSN), which is a structure composed of individuals and social interactions between them. It acts as a medium for spreading information, ideas or opinions among users through the effect of ``word-of-mouth''. For example, Facebook has over 1.52 billion daily active users and Twitter has 321 million monthly active users. The phenomenon of information diffusion in social networks has attracted the attention of scholars, businessmen and politicians because of its huge value in many aspects. These values can be used in many specific applications, such as virus marketing \cite{domingos2001mining} \cite{richardson2002mining}, network monitoring \cite{leskovec2007cost}, rumor blocking \cite{chen2019minimizing} \cite{guo2020multi}, and social recommendation \cite{ye2012exploring}. As a result, they jointly promote research on the Influence Maximization (IM) problem.
    
    The IM problem is the most important algorithmic problem in the research of information diffusion in social networks, which has been extensively studied for twenty years. Kempe \textit{et al.} \cite{kempe2003im} formally defined the IM problem as a combinatorial optimization problem in 2003, which aims to select a size-$k$ seed set such that the expected number of activated users (influence spread) can be maximized given a diffusion model such as Independent Cascade (IC) model and Linear Threshold (LT) model. This problem persists because of its inherent complexity. First, it has been proven that the IM problem is NP-hard \cite{kempe2003im}, thus there is no exact solution within polynomial running time. Second, due to the stochasticity of information diffusion process, it is \#P-hard to accurately compute the influence spread given a seed set under the commonly used diffusion models \cite{chen2010scalable_ic} \cite{chen2010scalable_lt}. Third, the IM problem depends on the given diffusion model and its corresponding parameters, but it is difficult to get it in practical applications. They have hindered the application of IM algorithms in large-scale social scenarios together.
    
    In the last two decades, the above three challenges have motivated a great deal of research to develop efficient algorithms for the IM problem. The existing work is mainly devoted to studying how to estimate the influence spread for the IM problem, and using the design techniques of heuristic algorithms or approximate algorithms to seek a balance between effectiveness, efficiency, and generalization ability. In this paper, heuristic algorithms and approximate algorithms are collectively referred to as ``traditional algorithms''. The heuristic algorithm is efficient and works well in some cases, but it has to be designed for a specific diffusion model, which is not universal, and there is no theoretical guarantee of the worst case. The time complexity of approximation algorithm cannot be further reduced on the premise of taking approximation into account, and it is difficult to meet the demand for algorithm performance due to the growing network scale. In addition, traditional algorithms do not have good generalization ability. In practical applications, frequent users' access leads to dynamic changes in social networks. When the network changes greatly, the traditional algorithm generally needs to search and solve again, or adjust the heuristic rules through trial and error to achieve better results. Therefore, these traditional methods have high computational cost and are difficult to be extended to large-scale, online, and real-time optimization problems, especially for variant IM problems.
    
    With the development of artificial intelligence technology, the method based on Machine Learning (ML) has surpassed traditional methods and achieved remarkable success in many fields. For example, Deep Reinforcement Learning (DRL) \cite{mnih2015human}, as an important branch of ML, achieves function approximation through Deep Neural Networks (DNNs), showing amazing learning and decision-making ability. It chooses an action according to the current state, and adjusts its own strategy according to the return brought by the action, which is called sequential decision-making. Based on that, there are several studies that recently begin to use DRL to learn a generalized policy for a certain combinatorial optimization problems on graphs \cite{khalil2017learning} \cite{nazari2018reinforcement} \cite{deudon2018learning} \cite{bengio2021machine}. The IM problem is a typical combinatorial optimization problem, that is, the optimal choice of decision variables in discrete space is naturally coupled with the action choice of reinforcement learning (RL). DRL has the characteristics of "offline training and online decision-making", which makes it possible to solve combinatorial optimization problems in online real-time manners. 
    
    Therefore, we can see that it is a feasible direction to use ML-based methods to solve the IM problem in social networks, so as to overcome the bottleneck of traditional algorithms. In this paper, we aim to make a comprehensive overview on the IM problem from an ML-based combinatorial optimization perspective. Here, we will focus on the following three aspects for detailed display.
    \begin{itemize}
        \item \textbf{The IM problem and traditional algorithms (Section \uppercase\expandafter{\romannumeral3}, \uppercase\expandafter{\romannumeral4}).} We first review the basic diffusion models, the definition of the IM problem, its properties, and computational complexity. Then, we introduce the classical greedy algorithm to study the IM problem, and review traditional algorithms, heuristic and approximation algorithms, according to the taxonomy of existing IM algorithms \cite{li2018influence}, including simulation-based, proxy-based, and sampling-based approaches. At last, we summarize the design bottlenecks and shortcomings of traditional algorithms, and point out the necessity of adopting ML-based methods at the same time.
        \item \textbf{DRL framework for the IM problem (Section \uppercase\expandafter{\romannumeral5}, \uppercase\expandafter{\romannumeral6}).} We first review the background and fundamental methods of adopting DRL to solve combinatorial optimization problems. These methods are mainly based on two categories: Pointer Networks and Graph Neural Networks (GNNs), each of which has different characteristics and applicability. Here, we mainly introduce the related concepts and methods that will be used in solving the IM problem later. Then, we review the current DRL methods to solve the IM problem, which relies on the techniques of graph embeddings and the DRL framework to solve the combinatorial optimization problems. Finally, we also summarize some ML-based methods from other niche technology stacks to address the IM problem.
        \item \textbf{The variants of IM problem and future research (Section \uppercase\expandafter{\romannumeral7}, \uppercase\expandafter{\romannumeral8}).} Based on the IM problem, there are a large number of variants that are adapted to different application scenarios, including the topic-aware IM problem \cite{chen2015online} \cite{tian2020deep}, the time-aware IM problem \cite{tong2020time} \cite{guo2021adaptive}, the location-aware IM problem \cite{li2014efficient} \cite{chen2020efficient}, the target-aware IM problem \cite{guo2019targeted} \cite{cai2020target}, and so on. Thus, we review the latest ML-based works that are used to solve different kinds of variant IM problems. Finally, we summarize the advantages and research opportunities of adopting ML-based methods to address the IM problem and the variants of IM problem in different social applications.
    \end{itemize}
    
\begin{figure}[!t]
	\centering
	\includegraphics[width=\columnwidth]{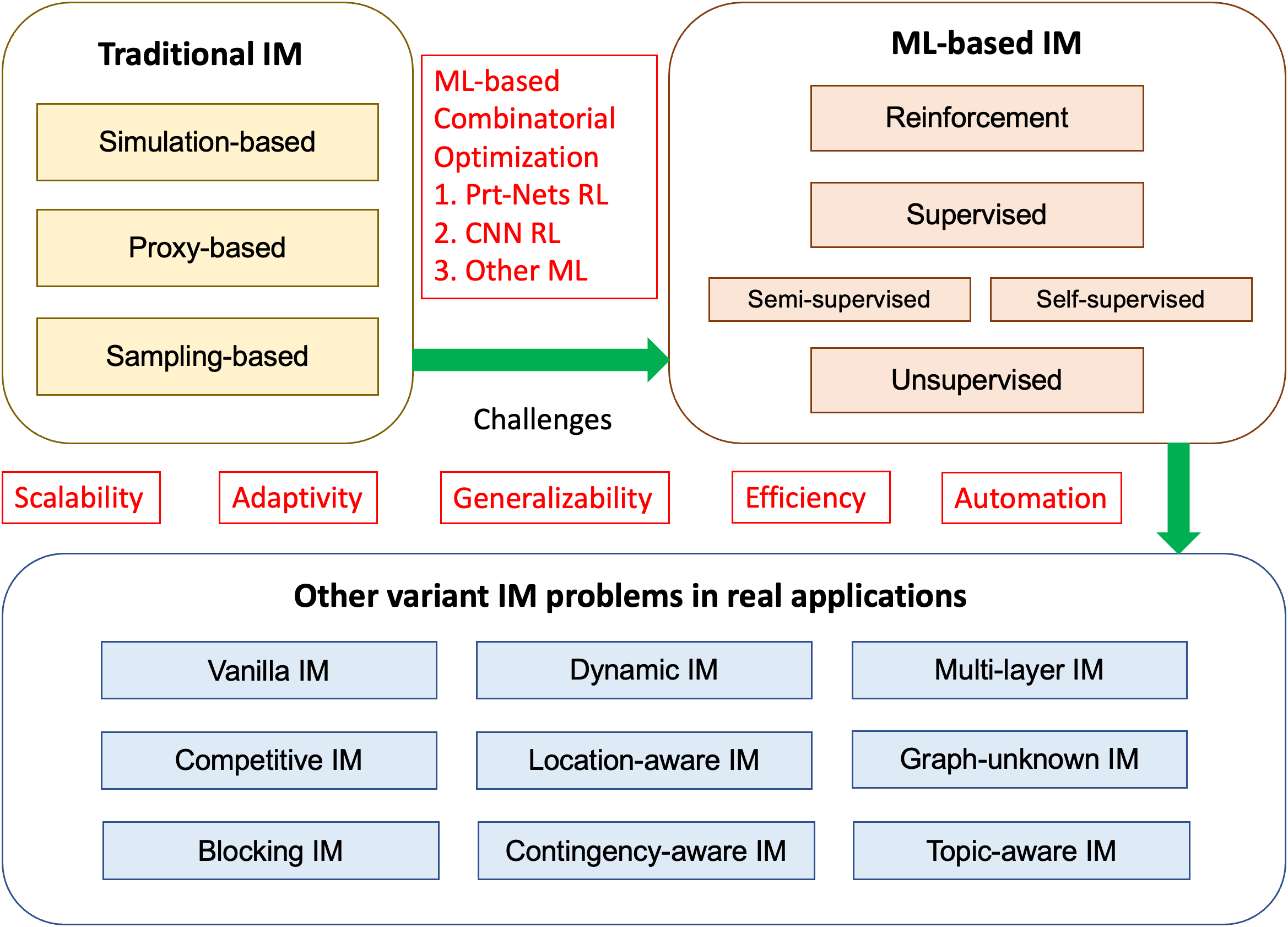}
	\caption{The architecture of this survey.}
	\label{fig1}
\end{figure}

    As a special combinatorial optimization problem, the IM problem has inherent complexity and vitality. To the best of our knowledge, we are the first to survey the IM problem from an ML-based combinatorial optimization perspective. The architecture of this paper is shown in Fig. \ref{fig1}. It involves a variety of emerging machine learning technologies such as reinforcement learning, graph learning, online learning, etc. Compared with traditional algorithms, it has great advantages in time complexity, generalization ability, and real-time performance. This paper not only overlooks the macro blueprint of this research field from a high-level perspective, but also explains it in detail. This is our main contribution.
    
\section{Related Works}
The IM problem has been a hot research topic in the field of network science, social computing, and combinatorial optimization for the past twenty years since Kempe \textit{et al.} \cite{kempe2003im} first defined the IM problem in 2003. There are a lot of existing surveys \cite{sun2011survey} \cite{guille2013information} \cite{chen2013information} \cite{tejaswi2016diffusion} \cite{arora2017debunking} \cite{li2018influence} about the IM problem and socal analysis. In \cite{sun2011survey} \cite{guille2013information} \cite{chen2013information} \cite{tejaswi2016diffusion} \cite{arora2017debunking}, they summarized some works of early IM algorithms, including a plenty of diffusion models, social influence analysis methods, and commonly used heuristic algorithms. Brogs \textit{et al.} \cite{borgs2014maximizing} proposed the technique of reverse influence sampling in 2014. Since then, the research of approximate algorithms for the IM problem has made breakthrough progress. In 2018, Li \textit{et al.} \cite{li2018influence} gave a comprehensive survey on the IM problem from an algorithmic perspective, which was the milestone of this area and covered almost all the traditional algorithms about the IM problem before this, including heuristic and approximation algorithms. They divided traditional algorithms into three categories: simulation-based, proxy-based, and sampling-based approaches. The introduction of traditional algorithms in this paper will continue this classification method and supplement some of the latest works.

After 2018, there have been some new surveys \cite{sumith2018influence} \cite{peng2018influence} \cite{jaouadi2019influence} \cite{banerjee2020survey} \cite{azaouzi2021new} \cite{aghaee2021survey} \cite{singh2021influence} \cite{zareie2021influence} \cite{chen2022influence} \cite{cai2022survey} \cite{taha2022identifying} on the IM problem. In \cite{sumith2018influence} \cite{peng2018influence} \cite{jaouadi2019influence} \cite{banerjee2020survey} \cite{azaouzi2021new} \cite{aghaee2021survey} \cite{singh2021influence} \cite{chen2022influence} \cite{taha2022identifying}, they were all based on previous heuristic or approximation algorithms. The only difference here was that some new elements have been added, including the latest methods, parallel or distributed computing, theoretical analysis and comparison, etc. In \cite{zareie2021influence} \cite{cai2022survey}, they considered two kinds of typical variant IM problems, behaviour-aware and location-aware IM problems. They summarized some updated research progress and challenges in this direction. 

Compared with the existing surveys, it is not difficult to find that they are all based on traditional algorithms and diffusion models. In this paper, we present a comprehensive review on the IM problem by introducing the latest techniques and approaches based on machine learning and give a fine-grained discussion and analysis on these new ML-based methods. This will be a pioneering and influential work.

\section{Influence Maximization}
In this section, we give some preliminary definitions about diffusion models, the IM problem and its hardness properties, and the greedy hill-climbing algorithm.

\subsection{Diffusion Models}
A social network can be denoted as a directed graph $G=(V, E)$, where $V$ with $|V|=n$ is the set of nodes in $G$ and $E$ with $|E|=m$ is the set of edges between nodes in graph G. A node $u\in V$ represents an individual or user in a social network, while an edge $(u,v)\in E$ indicates a certain type of social connection between user $u$ and $v$. For each edge $(u,v)\in E$, we say that $u$ is the in-neighbor of $v$ and $v$ is the out-neighbor of $u$. For each node $v\in V$, we denote by $N^-(v)$ the set of its in-neighbors and $N^+(v)$ the set of its out-neighbors. A node $u\in V$ is a seed node if it acts as the source of information diffusion at the beginning. The set of seed nodes is called seed set, denoted as $S$, $S\subseteq V$. The influence of the given seed set is based on the information diffusion process. Here, we consider that a user is active if she accepts (is activated by) the information cascade from her in-neighobors or she is selected as a seed. 

Then, we first define two commonly used diffusion model, Independent Cascade (IC) model and Linear Threshold (LT) model \cite{kempe2003im}, as follows in order to quantify the information diffusion across the social network. It can be represented by a discrete-time stochastic process, where we denote by $S_i\subseteq V$ the set of all active nodes at time step $t_i$. At time step $t_0$, all seed nodes in $S$ are activated, $S_0:=S$.
\begin{defn}[IC model]
    There is a diffusion probabiltiy $p_{uv}\in(0,1]$ associated with each edge $(u,v)\in E$. We set $S_i:=S_{i-1}$ at time step $t_i$ $(t\geq 1)$ first; then, for each node $u\in S_{i-1}\backslash S_{i-2}$ activated first at time step $t_{i-1}$, it have one chance to activate each of its inactive out-neighbor $v$ with probability $p_{uv}$. We add $v$ into $S_i$ if $u$ activates $v$ successfully at $t_i$. The influence diffusion terminates when no more inactive nodes can be activated.
\end{defn}
\begin{defn}[LT model]
    Each edge $(u,v)\in E$ has a weight $b_{uv}$, and each node $v\in V$ has a threshold $\theta_v$ sampled uniformly in $[0,1]$ and $\sum_{u\in N^-(v)}b_{uv}\leq 1$. We set $S_i:=S_{i-1}$ at time step $t_i$ $(t\geq 1)$ first; then, for each inactive node $v\in V\backslash S_{i-1}$, it can be activated if $\sum_{u\in S_{i-1}\cap N^-(v)}b_{uv}\geq\theta_v$. We add $v$ into $S_i$ if $v$ is activated successfully at $t_i$. The influence diffusion terminates when no more inactive nodes can be activated.
\end{defn}
\noindent
The diffusion probability $p_{uv}$ under the IC model can be regarded as the weight of edge directed from node $u$ to node $v$, which would distinctly affect the diffusion process. As a model parameter, the weight of edge should be learnt from the data ideally \cite{goyal2010learning} \cite{kutzkov2013strip} \cite{mathioudakis2011sparsification}. However, in practice, most of the existing works on the IM problem are still adopting pre-defined assignment for the sake of algorithm comparison \cite{arora2017debunking}. A commonly used weight assignment mechanism is called \textit{weighted cascade} (WC), which assigns $1/N^-(v)$ to weight $p_{uv}$. The weight $b_{uv}$ under the LT model is primarily assigned according to some heuristic mechanism, e.g. uniformly assignment as WC \cite{chen2010scalable_lt} \cite{galhotra2016holistic} or random assigning a value from $[0,1]$ before normalization \cite{chen2010scalable_lt} \cite{tang2014influence} \cite{tang2015influence}.

Then, Kempe \textit{et al.} \cite{kempe2003im} took the IC model and LT model as special cases, and summarized them to the Triggering model.
\begin{defn}[Triggering model]
Each node $v$ selects a triggering set $T_v$ randomly and independently according to a distribution $\mathcal{D}_v$ over the subsets of $N^-(v)$. We set $S_i:=S_{i-1}$ at time step $t_i$ $(t\geq 1)$ first; then, for each inactive node $v\in V\backslash S_{i-1}$, it can be activated if there is at least one node in $T_v$ activated in $t_{i-1}$. We add $v$ into $S_i$ if $v$ is activated successfully at $t_i$. The influence diffusion terminates when no more inactive nodes can be activated.
\end{defn}
\noindent
From above, a triggering model can be defined as $\Omega=(G,\mathcal{D})$, where $\mathcal{D}=\{\mathcal{D}_{v_1},\mathcal{D}_{v_2},\cdots\mathcal{D}_{v_n}\}$ is a set of distribution over the subsets of each $N^-(v_i)$. For each node $v\in V$, each node $u\in N^-(v)$ appears in $v$'s random triggering set $T_v$ with probability $p_{uv}$ independently under the IC model; and at most one node can appear in $T_v$, thus, for each node $u\in N^-(v)$, $T_v=\{u\}$ with probability $b_{uv}$ exclusively and $T_v=\emptyset$ with probability $1-\sum_{u\in N^-(v)}b_{uv}$ under the LT model.

\subsection{The definition of IM problem}
The IM problem is to find a seed set $S\subseteq V$ such that its influence spread $\sigma(S)$, can be maximized, which is the expected number of active nodes after the diffusion terminates. To mathematically define the influence spread, we first introduce a concept called ``realization'' or ``possible world''. Given a triggering model $\Omega=(G,\mathcal{D})$, Guo \textit{et al.} \cite{guo2021continuous} defined the realization as follows.
\begin{defn}[Realization]
Given a directed graph $G=(V,E)$ and triggering model $\Omega=(G,\mathcal{D})$, a realization $g=\{T_{v_1},T_{v_2},\cdots,T_{v_n}\}$ of $G$ is a set of triggering set sampled from distribution $\mathcal{D}$, denoted by $g\sim\Omega$. For each node $v\in V$, we have $T_{v}\sim\mathcal{D}_{v}$ respectively.
\end{defn}
If a node $u$ appears in $v$'s triggering set, $u\in T_v$, we say edge $(u,v)$ is live, or else edge $(u,v)$ is blocked. Thus, realization $g$ can be regarded as a subgraph of $G$, which is the remaining graph by removing these blocked edges. Let $\Pr[g|g\sim\Omega]$ be the probability of realization $g$ of $G$ sampled from distribution $\mathcal{D}$, that is,
\begin{equation}
\Pr[g|g\sim\Omega]=\prod\nolimits_{i=1}^{n}\Pr[T_{v_i}|T_{v_i}\sim\mathcal{D}_{v_i}]
\end{equation}
where $\Pr[T_{v_i}|T_{v_i}\sim\mathcal{D}_{v_i}]$ is the probability of $T_{v_i}$ sampled from $\mathcal{D}_{v_i}$. Under the IC-model, $\Pr[T_{v}|T_{v}\sim\mathcal{D}_{v}]=\prod_{u\in T_{v}}p_{uv}\prod_{u\in N^-(v)\backslash T_v}(1-p_{uv})$, and under the LT-model, $\Pr[T_{v}=\{u\}|T_{v}\sim\mathcal{D}_{v}]=b_{uv}$ for each $u\in N^-(v)$ and $\Pr[T_{v}=\emptyset|T_{v}\sim\mathcal{D}_{v}]=1-\sum_{u\in N^-(v)}b_{uv}$ deterministically.

The influence cascade on a realization becomes deterministic instead of stochastic process. As a result, the influence spread across the network can be regarded as the expected spread on all possible realizations. Now, the IM problem can be written in an expectation form as follows:
\begin{defn}[Influence Maximziation]\label{im}
	Givne a social graph $G=(V,E)$, a triggering model $\Omega=(G,\mathcal{D})$, and a budget $k$, the IM problem asks to find a seed set $S^\circ\subseteq V$, with at most $k$ nodes, that can maximize the expected influence spread across the graph, i.e.,
	\begin{equation}
	    S^\circ\in\arg\max_{|S|\leq k}\sigma_\Omega(S)=\mathbb{E}_{g\sim\Omega}[|I_g(S)|]=\sum\nolimits_{g}\Pr[g|g\sim\Omega]\cdot |I_g(S)|,
	\end{equation}
	where $I_g(S)$ contains all nodes can be reached from a node in $S$ by the live edges in the realization $g$.
\end{defn}

The IM problem is NP-hard under the IC model, LT model, and Triggering model \cite{kempe2003im}. The core idea of proofs is to reduce from the NP-complete Set Cover problem to the IM problem under the IC model and the NP-complete Vertex Cover problem to the IM problem under the LT model. The Triggering model is a generalized form of IC model and LT model, thus it easily has the same hardness by a simple induction.

\subsection{Greedy Algorithm and Hardness}
Due to the computational hardness of IM problem, the optimal solution cannot be exactly calculated out within polynomial time. Fortunately, the influence function $\sigma_\Omega(S)$ satisfies two properties, monotonicity and submodularity, which is helpful to design approximate algorithms. A lot of subsequent IM algorithms are based on these two properties and related optimization theory.
\begin{defn}[Monotonicity and Submodularity]
    Given a set function $f : 2^V\rightarrow\mathbb{R}_+$ and any two subsets $S$ and $T$ with $S\subseteq T\subseteq V$, we say it is monotone if $f(S)\leq f(T)$ and submodular if $f(S\cup\{v\})-f(S)\geq f(T\cup\{v\})-f(T)$.
\end{defn}
\noindent
The IM problem is NP-hard and the influence function $\sigma_\Omega(S)$ is monotone and submodular with respect to $S$ under the Triggering model.

\begin{algorithm}[h] 
	\caption{{Greedy $(G, \Omega, k)$}}
	\label{a1}
	$S\leftarrow\emptyset$\;
	\For{$i=1$ to $k$}{
	    $u^*\leftarrow\arg\max_{u\in V\backslash S}\{\sigma_\Omega(S\cup\{u\}-\sigma_\Omega(S)\}$\;
	    $S\leftarrow S\cup\{u^*\}$\;
	}
	\Return $S$
\end{algorithm}

Most of the traditional IM algorithms are derived from the greedy hill-climbing algorithm, which is shown in Algorithm \ref{a1}. Shown as Algorithm \ref{a1}, it initializes an empty seed set $S$, and iteratively select a node $u^*$ with the maximum marginal gain $\sigma_\Omega(u|S)=\sigma_\Omega(S\cup\{u\})-\sigma_\Omega(S)$ until there are $k$ nodes in the seed set. Nemhauser \textit{et al.} \cite{nemhauser1978analysis} \cite{fisher1978analysis} build a theoretical guarantee for maximizing a monotone and submodular function under different constraints in 1978. They point out that given a non-negative monotone and submodular function under the cardinality constraint, the greedy algorithm can guarantee a $(1-1/e)$ approximation under the value oracle. However, the value oracle cannot be satisfied with the IM problem because of the stochasticity of information diffusion. Chen \textit{et al.} \cite{chen2010scalable_ic} \cite{chen2010scalable_lt} present that given a seed set $S$, it is \#P-hard to compute the accurate influence spread $\sigma_\Omega(S)$ under the IC model and LT model. The most direct way to estimate the influence spread is to use Monte Carlo simulation, where we repeatedly simulate the diffusion process of a given seed set and take its average value as the final result. This estimation introduces an error term $\varepsilon$, and then the greedy algorithm can only return a $(1-1/e-\varepsilon)$ approximation within the $\Omega(kmn\cdot poly(1/\varepsilon))$ time complexity.

In the greedy algorithm, it uses Monte Carlo simulation to estimate the marginal gain for all nodes in each iteration, leading to unacceptable running time when applied in large-scale social networks. Therefore, a series of subsequent work focused on how to improve the running efficiency of IM algorithms. Their goals are to effectively estimate the influence spread, thus overcome the \#P-hardness. According to the theoretical guarantee of the worst case, the IM algorithms can be divided into heuristic algorithms and approximation algorithms. According to different design strategies of estimating the influence spread, the IM algorithms can be categorized into three classes \cite{li2018influence}, i.e., simulation-based, proxy-based, and sampling-based approaches. Among them, the simulation-based and sampling-based approaches are mainly based on approximation algorithm, but proxy-based approaches are mainly based on heuristic algorithm. The representative algorithms of these three classes are listed as follows, and we will introduce them in detail in the next section.
\begin{itemize}
   \item Simulation-based: CELF \cite{leskovec2007cost}, CELF++ \cite{goyal2011celf++}, UBLF \cite{zhou2015upper}, CGA \cite{wang2010community}, and SA \cite{jiang2011simulated}.
   \item Proxy-based: SP \cite{kimura2006tractable}, MIA/PMIA \cite{chen2010scalable_ic}, LDAG \cite{chen2010scalable_lt}, SIMPATH \cite{goyal2011simpath}, and EASYIM \cite{galhotra2016holistic}.
   \item Sampling-based: TIM/TIM+ \cite{tang2014influence}, IMM \cite{tang2015influence}, SSA/D-SSA \cite{nguyen2016stop}, OPIM-C \cite{tang2018online}, and HIST \cite{guo2020influence} \cite{guo2022influence}.
\end{itemize}

\section{Traditional Algorithms}
In this section, we introduce three kinds of traditional algorithms respectively, explain the difficulties faced by traditional algorithms, and supplement some of the latest works.

\subsection{Simulation-based Approach}
    The core idea of this category is to adopt Monte Carlo (MC) simulation to estimate the influence spread $\sigma_\Omega(S)$. Specifically, with the stochastic traits of the diffusion process, the activation behaviour between two neighboring nodes occurs with some probability, which leads to simulation being a straightforward approach for calculating $\sigma_\Omega(S)$. The seminal work on the simulation-based approach has been discussed in the last section. To estimate a marginal gain, it conducts $r$ rounds of MC simulation to get $r$ different realizations based on the distribution given by a diffusion model. Note that the estimation of the influence spread $\sigma_\Omega(S)$ becomes more and more accurate as the value of $r$ increases. Then the node $u^*$ with highest estimated marginal gain will be added to the seed set $S$. To make the error small enough, $r$ can be empirically set to a large number such as $r=10000$ \cite{kempe2003im}. Chen \textit{et al.} \cite{chen2013information} quantify the relationship between $r$ and error $\varepsilon$, and point out that the simulation-based greedy algorithm returns a $(1-1/e-\varepsilon)$ approximation if the number of MC simulation $r$ is given by $\Theta(k^2n\log(n^2k)/\varepsilon^2)$ under the triggering model, then the algorithm runs in time $O(k^3n^2m\log n/\varepsilon^3)$.
    
    The unacceptable running time of the simulation-based greedy algorithm leads to a lot of researches trying to reduce the number of MC simulation. Leskovec \textit{et al.} \cite{leskovec2007cost} propose the CELF algorithm in 2007, which reduces the time complexity on the premise of ensuring the approximate ratio. They devise a mechanism with early stopping to largely improve the efficiency through identifying and pruning insignificant nodes to reduce the usage of MC simulation, but it does not improve the worst-case time complexity. In the greedy algorithm, there are $k$ iterations, and in each iteration, it would choose one node with maximum marginal gain, which normally require to traversal over all nodes out of seed set $S$ to compute the marginal gain. However, CELF introduces an upper bound attached to each node whose value is its marginal gain in some past iteration. Thus, we do not need to compute the marginal gain if its upper bound is already smaller than the maximum of those visited nodes' marginal gain that have been computed in the current iteration. Specifically, in the first iteration, the upper bound of each node $u\in V$ is assigned to the $\sigma_\Omega(\{u\})$. During the iteration, the nodes out of seed set are traversed in descending order based on their upper bounds to compute the marginal gain. Here, it introduces an early stopping mechanism that the traversal within an iteration would be stopped if the maximum upper bound of all unvisited nodes is smaller than at least one marginal gain of the visited nodes. Once the early stopping triggered, the current iteration would be terminated, the upper bound of each visited node would be updated with the latest marginal gains. The idea behind is that the marginal gain of a node in the current iteration cannot be larger than its marginal gain in the previous iteration, and hence, maintains it while selecting seeds in subsequent iterations. This early stopping mechanism can greatly accelerate the greedy algorithm and achieve up to 700 times faster speed. Goyal \textit{et al.} \cite{goyal2011celf++} design the CELF++ algorithm to further reduce unnecessary MC simulation on the basis of CELF by including the marginal gain of the node that has the best marginal gain in the previous iterations. UBLF \cite{zhou2015upper} utilizes a quick method to estimate the upper bound of $\sigma_\Omega(\{u\})$ for all $u\in V$ by using matrix analysis, and prove it is much faster than CELF/CELF++. However, it can only be used under the IC model and LT model.
    
    In addition to reducing the number of MC simulation, there are some researches, such as CGA \cite{wang2010community} and SA \cite{jiang2011simulated}, aiming to reduce the MC complexity. Here, they design some heuristic rules to simplify the process of MC simulation, which cannot guarantee the approximation ratio. In summary, the simulation-based approach has the benefit of model generalization, which implies that it does not depend on diffusion models since MC simulation is pervasively adaptive for arbitrary diffusion models. However, the time complexity of MC simulation is exponentially high due to the existence of parameter $r$ even with acceleration techniques such as CELF/CELF++, which is inherently hard to be further improved and cannot be applied to large-scale social networks.
    
\subsection{Proxy-based Approach}
    The core idea of this category is to adopt proxy models to approximate the influence spread $\sigma_\Omega(S)$ given a seed set $S$, thereby avoiding time-consuming MC simulation. The general method is to simplify the information diffusion process and replace the real diffusion model with the proxy model to estimate the influence spread, so as to overcome its \#P-hardness and achieve the purpose of improving efficiency. Proxy models are usually designed according to some heuristic rules, and there are two typical strategies: (1) Simplifying the stochastic diffusion model to the deterministic model such that the influence spread can be exactly computed; and (2) Restricting the influenced scope of each node to the  adjacent hops and ignoring the further hops.
    
    Under the IC model, Kimura \textit{et al.} \cite{kimura2006tractable} put forward a Shortest-Path (SP) model, which only considered the shortest path from node $u$ to $v$ when computing the influence spread. Chen \textit{et al.} \cite{chen2010scalable_ic} propose a representative proxy model in 2010, MIA/PMIA model. The key idea of MIA is to reduce stochastic diffusion models into deterministic models by constructing special spanning trees of propagation for each node $v$ to avoid MC simulation and speed up computing efficiency. Specifically, for a node $v$, MIA first builds up an in-arborescence subgraph ending with $v$ and an out-arborescence subgraph starting from $v$ where the paths between any node $u$ and $v$, namely $(u,v)$ or $(v,u)$, are selected through Dijkstra SP algorithm with designed length metric inversely proportional to the likelihood of a path under the IC model. Some paths in the subgraphs are then pruned if their likelihood is less than a given threshold $\theta$, and the remaining subgraphs are kept for further computation of the influence spread. Thus, the nodes influenced by a node $u$ can be directly targeted if there exists paths. PMIA is an extension of MIA for more accurate influence estimation, updating the in-arborescence subgraph after adding a node into the seed set so that existing seeds will not block the influence of future seeds. Since the influence function under the MIA/PMIA model is monotone and submodular, with the greedy framework, the $(1-1/e)$ approximation ratio can be guaranteed under these proxy models. Afterwards, Kim \textit{et al.} \cite{kim2013scalable} continue to reduce the IC model similar to MIA model by the independent path algorithm, which prunes all influence paths with the probability less than a given threshold.
    
    Under the LT model, Chen \textit{et al.} \cite{chen2010scalable_lt} propose a pioneering proxy model in 2010, LDAG model. It uses directed acyclic graphs (DAGs) to reduce the stochastic diffusion models, which constructs DAG for each node by Dijkstra SP algorithm and prunes the paths with the probability less than the threshold $\theta$ in a similar way to MIA/PMIA, then estimates the influence based on those DAGs. Because of the NP-hardness of computing the optimal DAG for a node, sub-optimal DAGs are used and extra error may be introduced. Goyal \textit{et al.} \cite{goyal2011simpath} design a SIMPATH algorithm where the influence spread can be estimated by counting the total number of simple paths beginning from each node of the seed set. Unlike LDAG enumerating all simple paths, SIMPATH estimates the influence spread considering fewer paths by removing the paths of which the likelihood is less than the threshold $\theta$. Thus, it is usually faster than LDAG. Besides, EASYIM \cite{galhotra2016holistic} can be used for both IC model and LT model. It estimates the influence of a node $u$ by computing the number of the paths starting from $u$ with length less than $l$ and count overlapping paths respectively to improve the quality of estimation.
    
    In addition, Li \textit{et al.} \cite{li2018influence} point out that there is another proxy-based approach called ``influence ranking proxy'', such as IRIE \cite{jung2012irie}, GroupPR \cite{liu2014influence}, SPIN \cite{narayanam2010shapley}, and IMRank \cite{cheng2014imrank}. Actually, they transform the IM problem to another problem, and rank all nodes according to some index, which have deviated from IM algorithms based on combinatorial optimization. Thus, we will not repeat them here. In summary, the proxy-based approach can significantly reduce the computation overhead of estimating the influence spread and show empirical efficiency in many IM-based applications. However, it lacks theoretical guarantee (approximation ratio), which leads to extreme unstability in the worst case. Furthermore, the proxy model has a drawback. It has to be designed relying on a specific diffusion model, which cannot be generalized to the other diffusion models.
    
\subsection{Sampling-based Approach}
    The core idea of this category is to sample a number of realizations from a graph given a diffusion model, and then estimate the influence spread of any seed set based on these sampled realizations. It avoids iteratively running MC simulation in the greedy algorithm, and can significantly improve the time efficiency and ensure the theoretical guarantee at the same time. Regarding the advance of effectiveness, we only focus on reviewing the algorithms based on Reverse Influence Sampling (RIS) rather than Forward Influence Sampling (FIS). The representative algorithms based on FIS are NEWGREIC \cite{chen2009efficient}, STATICGREEDY \cite{cheng2013staticgreedy}, StaticGreedyDU \cite{cheng2013staticgreedy}, PRUNEDMC \cite{ohsaka2014fast}, and SKIM \cite{cohen2014sketch}. Even though FIS-based algorithms formed a pioneering sampling-based approach, it has been quickly replaced by RIS-based algorithms because of its limited improvement in time efficiency. Thus, there is no one cares about it now.
    
    In 2014, Borgs \textit{et al.} \cite{borgs2014maximizing} first propose the technique of RIS, which is a milestone in the development history of traditional IM algorithms. This technique is based on the concept of random reverse reachable (RR) set. A random RR set can be generated in following three steps:
    \begin{itemize}
        \item Uniformly select a node $v$ from $V$.
        \item Randomly sample a realization $g$ from $G$ under a given diffusion model $\Omega$.
        \item Generate RR set, denoted by$R$, by collecting the nodes that can reach $v$ through a live path in realization $g$.
     \end{itemize}
    For each node $u\in V$, the probability that it is contained in $R$ randomly generated by the $v$ equals the probability that $u$ can activate $v$. Given a seed set $S$ and a random RR set $R$ sampled by a diffusion model $\Omega$, we have
    \begin{equation}
        \sigma_\Omega(S)=n\cdot\mathbb{E}_R[\mathbb{I}(S\cap R)],
    \end{equation}
    where $\mathbb{I}(S\cap R)=1$ if $S\cap R\neq\emptyset$, else $\mathbb{I}(S\cap R)=0$. Thus, the general framework of RIS-based algorithms is to first generate a collection of random RR set $\mathcal{R}=\{R_1,R_2,\cdots,R_\theta\}$. Here, it estimates $\sigma_\Omega(S)$ by $n\cdot Cov_\mathcal{R}(S)/\theta$ where $Cov_\mathcal{R}(S)$ is the number of RR sets covered by the seed set $S$. The IM problem has been transformed to the maximum coverage problem, and then the greedy algorithm is used to select a seed set $S^*$ that covers the maximum number of RR sets.
    
    Obviously, the estimation will be more and more accurate with the increase of $\theta$.  In order to strike a balance between computation efficiency and estimation accuracy, the subsequent researches concentrate on how to determine the value of $\theta$ and how to bound the error and get a $(1-1/e-\varepsilon)$ approximation with a high probability. Borgs \textit{et al.} \cite{borgs2014maximizing} design a threshold-based method that keeps generating RR sets until the total number of edges examined during the generation process reaches a pre-defined threshold $\tau$. They prove that it returns a $(1-1/e-\varepsilon)$ approximate solution with $1-1/n$ probability in polynomial time when setting $\tau$ to $O(k(m+n)\cdot\log^2 n/\varepsilon^3)$. Tang \textit{et al.} \cite{tang2014influence} propose a TIM/TIM+ that follows the RIS to improve efficiency. They were the first to directly build the relationship between the number of RR sets $\theta$ and approximation loss, whose expected time complexity is $O((m+n)\cdot\log n/\varepsilon^2)$ when ensuring the same theoretical bound. Tang \textit{et al.} \cite{tang2015influence} \cite{chen2018issue} further improve the TIM/TIM+ and proposed an IMM algorithm. It adopts a martingale analysis to better estimate the lower bound of optimal value, and further scales down the number of RR sets while keeping the same worse-case guarantees. Nguyen \textit{et al.} \cite{nguyen2016stop} propose a SSA/D-SSA algorithm that iteratively doubles the number of RR sets and selects the seed set based on current collection of RR sets until satisfying the stopping condition. They claim that they achieve much better performance than previous methods, but their theoretical analysis is doubted by other researchers \cite{huang2017revisiting} \cite{nguyen2018revisiting}. In 2018, Tang \textit{et al.} \cite{tang2018online} change this pattern by adopting an online sampling method. They created an OPIM-C, which sampled two collections of random RR sets, one for selecting a seed set and the other one for verifying whether the seed set is qualified, until a $(1-1/e-\varepsilon)$ approximation can be provided. It keeps the same expected time complexity as IMM but achieves faster in practice because of the much smaller number of RR sets needed to be generated.
    
    The above-mentioned RIS-based algorithms may result in a large memory consumption because a large number of RR sets should be stored in the memory used for the seed set selection. To effectively reduce the memory consumption, Wang \textit{et al.} \cite{wang2016bring} propose a lazy sampling technique (BKRIS) that adopts the bottom-K minHash technique and selects the seed set without fully implementing all $\theta$ RR sets. Guo \textit{et al.} \cite{guo2020influence} \cite{guo2022influence} propose a SUBSIM framework to speed up the generation process of RR sets and create a Hit-and-Stop (HIST) algorithm to tackle the scalability issue in high influence networks by reducing the average size of random RR sets without losing the approximation guarantee, thus saving the memory consumption.
    
    In summary, the sampling-based approach has advantages over the other two approaches. It has a rigorous approximation guarantee and relatively low time complexity, which achieves the balance between theoretical guarantee and algorithm efficiency. It represents the highest level of traditional algorithms. Nevertheless, the sampling should be aligned with the given diffusion model. Unlike the proxy-based approach, it does not need to be specially designed according to the diffusion model, but does not mean that any diffusion model can easily give an explicit distribution so that the realization can be sampled. This is a bad thing. Next, RIS-based algorithms have to store a large number of RR sets, which is a waste of memory and storage resources. The setting of parameters in these algorithms heavily affects the approximation and time complexity, which shows inadequate generalization ability and can barely adjust to dynamic social networks.
    
\section{Deep Reinforcement Learning for Combinatorial Optimization} \label{DRL for CO}
    As mentioned in the previous section, there is no traditional algorithm that can perfectly solve the IM problem. For heuristic algorithms, they have to be designed according to given diffusion models and have no theoretical guarantee. For approximation algorithms, their scalability has reached the limit and time efficiency cannot be further optimized. No matter which traditional algorithm we use, it has a poor generalization ability. That is to say, when the network topology changes, the algorithm must be re-run to keep the validity of the seed set. In recent years, a new technique of solving combinatorial optimization problems by deep reinforcement learning (DRL) has emerged, which can effectively address the drawbacks of traditional algorithms. The IM problem is a special kind of combinatorial optimization problems. Thus, before introducing machine learning (ML) based IM algorithms, we first survey some latest development of DRL for combinatorial optimization. We will focus on the concepts, technologies, and methods related to ML-based IM algorithms.
    
    The Combinatorial Optimization (CO) problem is a sub-field of mathematical optimization problem, which is defined in the discrete space. The CO problem can be defined by a standard form, which is mathematically formulated as follows:
    \begin{align}
		\min\quad&F(x)\\
		&\text{s.t. }G(x)\geq 0, x\in D,
	\end{align}
    where $x$ is the decision variable, $F(x)$ is the objective function, $G(x)$ is the constraint condition, and $D$ is a discrete space. Typical CO problems include Traveling Salesman Problem (TSP), Vehicle Routing Problem (VRP), Job Scheduling, Knapsack, Minimum Vertex Cover (MVC), Minimum Dominating Problem (MDoP), and so on.
    
    Similar to traditional algorithms for the IM problem, traditional algorithms for CO problems consist of exact, approximate, and heuristic approaches. The exact approach primarily adopts the idea of "divide and conquer", to dividing the problem into smaller sub-problems and solve them recursively to finally achieve the optimal solution, such as Branch-and-Bound \cite{lawler1966branch} and Dynamic Programming \cite{bertsekas2011dynamic}. The approximate approach aims at finding local optimal solution for higher efficiency, such as greedy algorithm and local search algorithm, which can provide theoretical guarantees for the quality of solutions. The heuristic algorithm, such as stimulated annealing algorithm \cite{van1992job} and evolutionary algorithms \cite{vikhar2016evolutionary}, can largely accelerate the computation by using specialized domain knowledge and involving trial-and-error as well as tuning. However, all of these approaches suffer from computational inefficiency for large-scale problems. In addition, for more and more applications that require high real-time performance, once the situation of problem changes, it is necessary to recalculate the solution, which would inevitably produce too expensive calculation costs. Thus, we need to find other efficient and effective solutions.
    
    A natural question is whether a ML-based algorithm can have generalization ability such that the algorithm of a CO problem learned from part of instances can be generalized to unseen instances with the same distribution. To answer this question, a lot of researchers have made efforts to it. Using ML-based methods to solve CO problems can be traced back to the Hopfield network \cite{hopfield1985neural} proposed in 1985. It can be used to solve the TSP \cite{smith1999neural} but only for those small-scale single instances. With the  development of hardware performance and deep learning, a lot of researchers began to study how to use ML-based methods to solve generalized NP-hard CO problems \cite{vinyals2015pointer} \cite{bello2016neural} \cite{khalil2017learning} \cite{kool2018attention} \cite{selsam2018learning} \cite{li2018combinatorial} \cite{nazari2018reinforcement} \cite{prates2019learning} \cite{lemos2019graph} \cite{ma2019combinatorial} \cite{manchanda2020gcomb} \cite{cappart2021combining} \cite{yuan2022rl} \cite{gunarathna2022solving}. They mainly rely on DRL because it can learn a policy to guide a series of decision making actions under different circumstances in order to gain a maximum accumulated benefit, which aligns with the characteristics of CO problems like TSP, Knapsack, IM, etc.
    
    According to the coding strategies of problem instances, the architecture can be roughly divided into two categories: one based on Pointer Networks (Ptr-Nets) \cite{vinyals2015pointer} and the other based on Graph Neural Networks (GNNs) \cite{khalil2017learning}. They form two kinds of end-to-end framework: "Ptr-Nets based RL" and "GNNs based RL". End-to-end framework takes a given problem instance as input, and can directly output the solution without searching. Its problem embedding and solving are integrated, whose training can be finished together. Once the training is completed, the problem instances with the same distribution characteristic can be directly solved. Therefore, it has the advantages of fast solution speed and strong generalization ability. We will introduce them separately. Here, we assume that readers have basic ML knowledge \cite{goodfellow2016deep} \cite{murphy2022probabilistic}.
    
    
\begin{figure}[!t]
	\centering
	\includegraphics[width=\columnwidth]{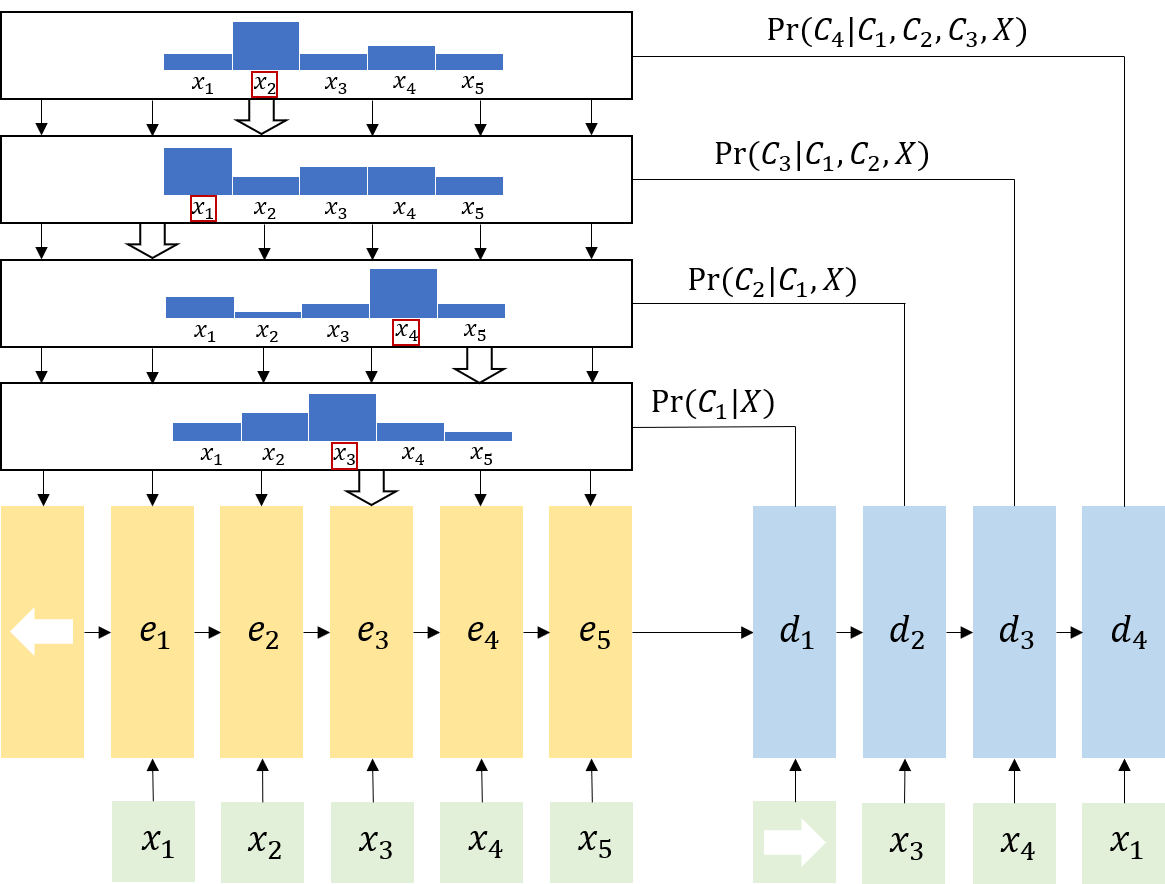}
	\caption{The architecture of Pointer Networks \cite{vinyals2015pointer}, where the yellow network is encoder and blue network is decoder. They interact through the attention mechanism.}
	\label{fig2}
\end{figure}
    
\subsection{Ptr-Nets Based RL Framework}
    The Prt-Net is a pioneering model, which learns from the sequence-to-sequence (seq2seq) model \cite{sutskever2014sequence} in the machine translation to solve CO problems, that is, encode the input sequence of problem instance by an encoder, then compute the selection probability of each node by a decoder and attention mechanism \cite{bahdanau2014neural}, and gradually select the nodes in an auto-regressive way until a complete solution. Its architecture is shown in Fig. \ref{fig2}, where the yellow network is encoder and the blue network is decoder. The encoder and decoder are implemented by a kind of Recurrent Neural Networks, such as RNN, LSTM, and GRU. It first converts the input sequence to a code and feeds it into the decoder, which generates an output according to content-based attention mechanism over input sequence.
    
    For convenience, we define the encoder and decoder hidden states as $(e_1,\cdots,e_n)$ and $(d_1,\cdots,d_m)$, respectively. The input is denoted by $X$ and the $i^{\text{th}}$ outcome is denoted by $C_i$. In the $i^{\text{th}}$ decoding, the RNN reads the hidden state of previous layer $d_{t-1}$ and outputs its hidden layer $d_t$. Based on $d_t$ and encoder hidden state, the posterior probability of the $i^{\text{th}}$ outcome can be modeled as follows:
    \begin{align}
        &u_{j}^{i} = v^T tanh(W_1 e_j + W_2 d_i),\quad j\in(1,\cdots,n),\\
        &\Pr(C_i|C_1,\cdots,C_{i-1},X) = \text{softmax}(u^{i}),
    \end{align}
    where softmax normalizes each component of the vector $u^i$ (of length $n$) to be an output distribution over the dictionary of inputs. The $v$, $W_1$, and $W_1$ are learnable (trainable) parameters of the decoder. Taking TSP for example, the input is the sequence of city coordinates, and the output is the solution of city indexes in order.
            
    The Ptr-Net can be trained through supervised approach or DRL approach. But due to the lack of labeled data (i.e. optimal routes for TSP) in practice, the current studies tend to train the model parameters such as $v$, $W_1$, and $W_1$ through DRL. The DRL formulates the optimizing objective as reward and learns policy to maximize the accumulated reward by sampling different events with corresponding instantaneous reward. The Value-based and policy-based algorithms are two mainstream approaches of DRL. Since the policy-based DRL approach models its stochastic policy considering state $S$ and action $A$ as $P_\theta(A|S)=\prod_{t=1}^{n}P_\theta(A_t|S,A_{1:t-1})$, which coincides with the mechanism given by Ptr-Nets. Policy-based DRL algorithms that take the total feedback as the standard and adopt the round update model, especially REINFORCE algorithm, are usually exploited to train Ptr-Nets.
            
    Vinyals \textit{et al.} \cite{vinyals2015pointer} propose the Ptr-Net in 2015 to solve three challenging geometric CO problems with supervised training, Planar Convex Hulls, Delaunay Triangulations, and Planar TSP. The supervised training requires a large number of labeled data, such as $(\textit{city coordinates}, \textit{optimal route})$ in TSP, to train the network parameters, which is hard to obtain. Also, supervised training inherently limits the quality of solutions to the training data, too dependent on the quality of labels. To address them, Bello \textit{et al.} \cite{bello2016neural} employ DRL framework to train the Ptr-Net on CO problems. They used REINFORCE algorithm to learn the parameters, taking each instance of the CO problem as a training sample and formulating the optimizing objective function as reward function in DRL. Actor-critic algorithm is introduced to restrain variance during training. The results outperform the Ptr-Net with supervised training \cite{vinyals2015pointer} and it achieves near-optimal solution to TSP as well as optimal solution to the knapsack problem. Motivated by Bello's work, Nazari \textit{et al.} \cite{nazari2018reinforcement} extend the Ptr-Net based RL to more complicated VRP, which take into account the static elements (customers' coordinates) and dynamic elements (customers' demands). By ignoring the order information of customers and replacing the LSTM of the input layer in the encoder, which would not impact the solution and greatly reduce the computation overhead. The results show comparable performance with Bello's work on TSP and superior to classic heuristics on VRP.
            
    With the development of attention mechanism, Transformer \cite{vaswani2017attention} adopts a multi-head attention mechanism instead of RNN/LSTM and enables models to extract deeper characteristics of data. With the tremendous success of Transformer in Natural Language Processing, many studies have begun to integrate Transformer into Ptr-Nets based RL to solve CO problems. Deudon \textit{et al.} \cite{deudon2018learning} upgrade the conventional Ptr-Net by using Transformer's multi-head attention structure to output the feature vector of nodes in encoder part, which effectively reduces the model complexity. They use this upgraded model to compute a preliminary result and then employ a simple 2OPT local search on it to compute eventual result, which shows improvement on the quality of solution. Kool \textit{et al.} \cite{kool2018attention} not only integrate the multi-head attention idea given by Transformer in the encoder part of Ptr-Nets but also adopt self-attention mechanism given by Transformer to compute the attention weights in the decoder part for better performance. Furthermore, they devise a Rollout baseline to substitute for the critic neural network in Actor-Critic architecture to accelerate the training efficiency and outperform the state-of-the-art (SOTA) end-to-end methods including those mentioned above. Ma \textit{et al.} \cite{ma2019combinatorial} propose a Graph Pointer Network (GPN) to solve TSP with time window, which is trained via a kind of hierarchical reinforcement learning (HRL). The GPN builds upon Ptr-Net by placing a graph embedding layer on the input, which captures relationships between nodes and generates graph embedding vectors as part of the input to decoder. They train the GPN through HRL to learn a hierarchical policy to find an optimal city permutation under constraints for TSP. With the graph embedding, the vector context information is merged so that the model trained on TSP 50/100 can generalize well to larger-scale TSP 500/1000 problems.
    
    However, none of the above methods reach the solution obtained by professional solvers in the TSP problem over 50 nodes. The scalability and generalization ability of this method has a big bottleneck needed to be solved. The Prt-Net is mainly used to solve CO problems with sequential characteristics such as TSP and VRP, where the solution of this kind of problems depends on the order of nodes. Thus, it is suitable for solving sequential CO problems because of using attention mechanism to construct solutions in an auto-regressive way.
            
\subsection{GNNs Based RL Framework}
    As we all know, the graphs has a complex structures, which can contain rich information about potential relationships between nodes \cite{barabasi2013network}. How to mine useful information from a large-scale graph has aroused people's attention. Graph Neural Networks (GNNs) is a new branch of methods that utilizes deep learning techniques to effectively analyze data with graph structures, which can abstract deep information from graph data. The key idea of GNNs is to encode the graph structure using a neural network model. 
    
    Generalization ability is a great challenge for traditional algorithms. Network encoding techniques can improve generalization ability by embedding a node $v$'s attributes as well as its associated local and global information in a graph $G$ into a $p$-dimensional representation vector $\mu_v$. We first introduce several typical methods of graph network encoding commonly associated with RL for solving the CO problem, then briefly review the studies in this area.
        
    \subsubsection{Random Walk Based Embedding} 
    \textit{DeepWalk} \cite{perozzi2014deepwalk} is proposed by Perozzi \textit{et al.} to learn representation vectors for a graph such as social networks and biological networks. It adopts random walks to extract information from a network, which samples a number of short trajectories that can be produced by random walks as sequential data with connectivity information. Arriving at a node $v$ in a walk, the possibility of choosing its neighboring node $u$ is proportional to the normalized weight of edge $(v,u)$. Then they use the language model \textit{SkipGram} \cite{mikolov2013efficient} to process those trajectory sequences and learn embeddings of nodes, similar to learning the word embedding from a given language dataset. \textit{node2vec} \cite{grover2016node2vec} extends \textit{DeepWalk}'s sampling method by introducing a biased random walk. Given current node $v$ during a walk, the possibility of visiting its neighboring node $u$ is updated by multiplying a coefficient $\alpha$ and the weight of edge $(v,u)$ on the basis of \textit{DeepWalk}. The coefficient $\alpha$ is related to two super parameters $p$ and $q$ which help to strike a balance between homophily and structural equivalence in node representation. This has been proved to be more effective.
    
    Both \textit{DeepWalk} and \textit{node2vec} are representation learning for homogeneous networks where only singular type of nodes and relationships exist. However, for ubiquitous heterogeneous networks in nature, especially social networks, it brings unique challenges resulting from diverse types of nodes and edges. Dong \textit{et al.} \cite{dong2017metapath2vec} develop two frameworks, \textit{metapath2vec} and its extension \textit{metapath2vec++}, to learn the representation for heterogeneous network. \textit{metapath2vec} aims at retaining information of structures and semantics of heterogeneous networks by adopting metapath-based \cite{sun2012mining} random walks to retrieve heterogeneous neighborhood trajectories and then use \textit{SkipGram} to generate node embeddings, which are expected to encode features of network structures and semantics. Moreover, \textit{metapath2vec++} integrates a heterogeneous negative sampling method on the basis of \textit{metapath2vec} to conduct simultaneous modeling of structures and semantics for efficient embedding. Extensive experiments show superior competence to the SOTA methods including \textit{DeepWalk} and \textit{node2vec} on heterogeneous networks.
    
    \subsubsection{Recurrent GNNs} 
    Dai \textit{et al.} \cite{dai2016discriminative} propose a \textit{structure2vec} for embedding networks, which has been widely used in subsequent social applications. It is inspired by graphical model inference algorithms, where node-specific features or attributes $x_v$ are aggregated recursively according to $G$’s graph topology. Specifically, given a partial solution $S$ which is being updated at each step of recursion, in $(i+1)^{\text{th}}$ iteration, \textit{structure2vec} calculates each node $v$'s temporary representation considering its features or attributes $x_v$, all its neighbors' representations in $i^{\text{th}}$ iteration $\{\mu_{u}^{(i)}\}_{u\in \mathcal{N}(v)}$, and the connectivity information between $v$ and all its neighbors in $i^{\text{th}}$ iteration $\{w(v,u)\}_{u\in \mathcal{N}(v)}$. One variant of the \textit{structure2vec} architecture can be formulated as follows:
    \begin{equation} \label{s2v}
        \mu_{v}^{(i+1)} \leftarrow F(x_v + \{\mu_{u}^{(i)}\}_{u\in \mathcal{N}(v)}, \{w(v,u)\}_{u\in \mathcal{N}(v)};\Theta),
    \end{equation}
    where $\mathcal{N}(v)$ is the neighbor set of node $v$ in graph $G$, $F$ is a generic nonlinear mapping to the current embedding such as a neural network or kernel function, $\Theta$ is the set of function parameters of $F$, and $\mu_{v}^{(0)}$ is initialized as 0 for all $v\in V$. The update rule shown as Eqn. (\ref{s2v}) indicates that the embedding is carried out based on the graph topology. After $I$ iterations, the embedding $\mu_{v}^{(I)}$ will definitely contain information about all neighbors within the $I$-hop neighborhood. The parameter set $\Theta$ can be learned through either supervised learning or reinforcement learning, the latter of which fits cases where the training labels are lacking.
            
    \subsubsection{Graph Convolutional Networks} The great success achieved by convolutional neural networks (CNNs) in structured data, such as feature extraction from pictures, motivates the appearance of Graph Convolutional Networks (GCNs) dealing with unstructured graph data. Similar to CNNs, GCNs define filters to do convolution operation with a graph signal $\vec{x}\in \mathbb{R}^n$ (a scalar for every node) yet in spectral domain in order to neglect irrespective features \cite{bruna2014spectral}. Given a graph $G$, the graph convolution operation, denoted by $*$, of two graph signals $\vec{x}_1$ and $\vec{x}_2$ is defined formally as follows:
    \begin{equation}\label{gconv}
        \vec{x}_1*\vec{x}_2 = \vec{U}((\vec{U}^{T}\vec{x}_1) \odot (\vec{U}^{T}\vec{x}_2)),
    \end{equation}
    where $\vec{U}$ is the matrix of eigenvectors of Laplacian matrix $\vec{L}$ of $G$ and $\odot$ denotes Hadamard product. The normalized Laplacian matrix of an undirected graph is defined as $\vec{L}=\vec{I}_n-\vec{D}^{-\frac{1}{2}}\vec{A}\vec{D}^{-\frac{1}{2}}$, where $\vec{A}$ is the adjacent matrix of $G$ and $\vec{D}\in \mathbb{R}^{n\times n}$ is a diagonal matrix with $\vec{D}(i,i)=\sum_i\vec{A}(i,j)$. With the semi-definite property, the normalized Laplacian matrix can be factored as $\vec{L}=\vec{U}\Lambda \vec{U}^{T}$, with a diagonal matrix of its eigenvalues $\Lambda$. Multiplying $\vec{U}^{T}$ transforms the graph signals $\vec{x}_1$ and $\vec{x}_2$ into the spectral domain (i.e. the graph Fourier transform), while multiplying $\vec{U}$ performs the inverse transform \cite{zhang2020deep}. Let us denote a filter $g_\theta = \text{diag}(\theta)$ with a parameter $\theta\in \mathbb{R}^n$ in the spectral domain. According to Eq. (\ref{gconv}), the output graph signal $\vec{x}'$ after filter $g_\theta$ can be obtained as follows:
    \begin{equation}\label{filter}
        \vec{x}' = g_\theta * \vec{x} = \vec{U} g_\theta \vec{U}^{T} \vec{x}.
    \end{equation}
    We can regard $g_\theta$ as a function of the eigenvalues of $\vec{L}$, i.e. $g_\theta(\Lambda)$. Due to the computational complexity of the eigenvector matrix $\vec{U}$ which is $\mathcal{O}(n^2)$, prohibitively expensive for large graphs, ChebNet \cite{defferrard2016convolutional} is proposed to simplify the computation process by adopting Chebyshev polynomials to approximate to $g_\theta(\Lambda)$. Kipf \textit{et al.}  \cite{welling2017semi} propose an algorithm which introduces a first-order approximation of ChebNet and improves the Eq. (\ref{filter}) as follows:
    \begin{equation}\label{Kipffilter}
        \vec{x}' = g_\theta * \vec{x} = \theta_0\vec{x}-\theta_1\vec{D}^{-\frac{1}{2}}\vec{A}\vec{D}^{-\frac{1}{2}}\vec{x}.
    \end{equation}
    Several improvements are further imposed by Kipf's work \cite{welling2017semi}. To alleviate overfitting, they reduce the number of parameters by assuming $\theta_0=\theta_1=\theta$ as follows:
    \begin{equation}
        \vec{x}' = g_\theta * \vec{x} = \theta(\vec{I}-\vec{D}^{-\frac{1}{2}}\vec{A}\vec{D}^{-\frac{1}{2}})\vec{x}.
    \end{equation}
    They generalize the model to a multi-channel signal $\vec{X}\in \mathbb{R}^n\times C$ with $C$ input channels (i.e. a $C$-dimensional feature vector for every node). To address the problem of numerical instability caused by using term $ \vec{I}-\vec{D}^{-\frac{1}{2}}\vec{A}\vec{D}^{-\frac{1}{2}}$, they applies a normalization trick of using $\tilde{\vec{D}}^{-\frac{1}{2}}\tilde{\vec{A}}\tilde{\vec{D}}^{-\frac{1}{2}}$ instead. This is the basic framework of GCNs, which is a prototype of many other GNN-based models.
            
    When processing graphs, algorithms should consider the neighbors of a node for aggregating information. As the number of neighbors of a node can be enormous in large graphs which are pervasive nowadays such as social networks, taking into account the full size of a node's neighborhood might lead to computational inefficiency. GraphSAGE \cite{hamilton2017inductive} addresses this problem by sampling a fixed number of neighbors for each node in order to reduce the time complexity. The graph convolution operation is performed as follows:
    \begin{align}
        &\vec{h}^{k}_{S_{\vec{N}(v)}} \leftarrow \text{AGGREGATE}_k\left (\left \{\vec{h}^{k-1}_u,\forall u\in S_{\mathcal{N}(v)}\right \}\right ),\\
        &\vec{h}^{k}_v \leftarrow \sigma \left (\vec{W}^k \cdot \left [\vec{h}^{k-1}_v,\vec{h}^{k}_{S_{\vec{N}(v)}} \right ]\right),
    \end{align}
    where $\vec{h}^{k}_v$ denotes the hidden feature vector of $v$ in the $k^{\text{th}}$ layer, $S_{\mathcal{N}(v)}$ denotes a random sample of the node $v$’s neighbors, $\text{AGGREGATE}(\cdot)$ denotes the aggregator function, $\sigma(\cdot)$ denotes a nonlinear activation function, and $[\cdot,\cdot]$ denotes the concatenation operation. The aggregation function should be invariant to the permutation of nodes, such as element-wise mean, sum, and max-pooling. 
    
    In the aforementioned GNNs, the models assume equal or manually defined weights for the neighbors of a node when aggregating them. The authors of Graph Attention Network (GAT) \cite{velivckovic2018graph} claim that the weights of neighbors may vary largely in practice. For example, different neighbors have different influence on a user. Thus the weights should be learned from data. GAT introduces attention mechanism into convolution-based GNNs in order to learn the corresponding weights between every two nodes. The convolutional operation of GAT \cite{wu2020comprehensive} is defined as follows:
    \begin{equation}
        \vec{h}^{k}_v = \sigma \left (\sum\nolimits_{u\in \mathcal{N}(v)\cup\{v\}} \alpha_{vu}^k \vec{W}^k \vec{h}_{u}^{k-1} \right),
    \end{equation}
    where $\vec{h}^{0}_v=\vec{x}_v$, $\sigma(\cdot)$ denotes a nonlinear activation function, and $\alpha_{vu}^k$ denotes the attention weight between node $v$ and its neighbor $u$ which is computed by the self-attention mechanism of the model.
            
    \subsubsection{Graph Autoencoders} Graph Autoencoder (GAEs) is an unsupervised learning method that encodes the node or graph into a hidden vector space through reconstructing the graphs' structural information, such as adjacent matrices, according to the hidden vector representation. Structural Deep Network Embedding (SDNE) \cite{wang2016structural} proposes a stacked autoencoder architecture, which learn effective embedding by tailoring the loss function, so as to consider the first-order proximity and the second-order proximity with two terms. The first term is designed for the first-order proximity by giving penalty for the distance of embedded vectors of two adjacent nodes, and the second term is designed for the second-order proximity by giving penalty for the difference between the input and reconstructed output of the adjacent matrix, where it assigns different weights to zero and non-zero elements.
    
    Here, we only give a simple review for GNNs that will be used in subsequent GNN-based RL frameworks and GNN-based IM algorithms. To learn more about GNNs, please read the latest review for deep learning on graphs \cite{zhang2020deep}. Next, we begin to introduce GNN-based RL framework.
            
    Dai \textit{et al.} \cite{khalil2017learning} propose a S2V-DQN framework that explores to solve the CO problem using GNNs and DRL. They utilize \textit{structure2vec} \cite{dai2016discriminative} as the encoder to embed node information of graph into latent node representation (i.e. embedding), and the Q-network in Deep Q-learning (DQN) algorithm takes the partial solution with latent node representation to predict the Q-value of each new node. The Q-value can be designed as the marginal gain of a node, which is normally implemented by a deep neural network having strong approximation ability. Then based on greedy policy, the model appends one node with largest Q-value to the partial solution set iteratively. Once the partial solution gets updated, the embedding of all nodes has to be re-generated. The model is trained on 50-100 nodes MVC, MAXCUT, and TSP, then gets tested on these problems with 1200 nodes, resulting in near-optimum and better generalization for graph out of distribution. Li \textit{et al.} \cite{li2018combinatorial} propose a GCN-TREESEARCH framework that adopts GCNs to directly estimate the probability of all nodes being selected, and based on the probability estimation, they exploit the search tree to construct a feasible solution. This method is suitable for node selection problems, regardless of the order of node selection, such as Maximal Independent Set (MIS), Maximal Clique (MC), and Satisfiability. Manchanda \textit{et al.} \cite{manchanda2020gcomb} propose a GCOMB framework, which employs a convolution-based GNN named GraphSAGE \cite{hamilton2017inductive} to predict the quality of a node and utilizes Q-learning of DRL techniques with importance sampling. It uses the GraphSAGE to prune poor nodes and learn embedding of good nodes in a supervised manner, and in turn Q-learning can only focus on good nodes for efficiency concern. All these methods \cite{li2018combinatorial} \cite{manchanda2020gcomb} give a better performance than the original one \cite{khalil2017learning}. 
    
    Besides, there are a lot of researchers using the above-mentioned GNN-based RL frameworks to solve CO problems in realistic applications. Qiu \textit{et al.} \cite{qiu2019dynamic} propose a multi-agent DRL approach with convolution-based GNNs to address the road tolling problem in a transportation network. They adopt an edge-based convolution-based GNNs to extract the spatio-temporal correlations of the road network state features where both the feature matrix and adjacency matrix in this problem are defined on edges instead of nodes. The model is trained using multi-agent DRL by decomposing the state and action spaces into sub-spaces and solving the problem in a divide-and-conquer manner. Mao \textit{et al.} \cite{mao2019learning} propose a model called \textit{Decima} to address the scheduling problem in data processing clusters. The \textit{Decima} introduces directed acyclic graphs (DAGs) to represent the graph of scheduling jobs and their dependencies, and it combines convolution-based GNN and policy gradient algorithm REINFORCE to embed the nodes and train the whole model. Ou \textit{et al.} \cite{ou2021active} address recurrent disease prevention on a social network by using GCNs and a two-level hierarchical DQN framework. Other applications can be seen in Virtual Network Function (VNF) placement \cite{sun2020combining}, adversarial attacks \cite{sun2020adversarial}, connected autonomous vehicles \cite{chen2021graph}, VNF forward graph placement \cite{xie2021virtualized}, route optimization \cite{almasan2022deep}, and mobile crowdsourcing \cite{xu2022adaptive}. They are all cases of successful application of ML method on graph-based problems.
    
    In summary, GNNs can learn the effective representation of graphs by their powerful learning ability, capturing the inner local and global features. The resulting embeddings can be utilized for future tasks such as classification or decision making. In recent years, GNN-based RL framework shows big potential in solving CO problems, especially for node selection problems such as MVC and MIS. On the one hand, nodes can still be selected step by step by an auto-regressive method \cite{khalil2017learning}. On the other hand, the selection probability of nodes can be directly estimated according to their features and graph topology, and it can use a searching technique to construct the feasible solution \cite{joshi2019efficient}.
    
\section{ML-Based Influence Maximization Algorihtms} \label{DRL for IM}
    As shown in Section \uppercase\expandafter{\romannumeral3} and \uppercase\expandafter{\romannumeral4}, the IM problem is a branch of CO problems, which is a typical node selection problem. Traditional algorithms for solving the IM problem usually require expertise knowledge and suffer from high computational cost. In addition, it lacks generalization ability for IM algorithms used in evolving graph or other variant problems. ML-based methods using deep learning techniques have the potential for solving these drawbacks. There are pioneering studies \cite{lin2015learning} \cite{ali2018boosting} combining RL with the IM problem. However, they do not model the influence estimation as an RL problem or node influence as a learning task. Thus, many subsequent researches \cite{li2019disco} \cite{manchanda2020gcomb} \cite{wang2021reinforcement} take advantage of graph embedding methods to learn representation of nodes or networks which captures the underlying association features for further use and contributes to generalization ability.
    
    In this section, we begin to review the deep learning techniques related to the vanilla IM problem, which is strictly defined by Definition \ref{im}, including graph embedding, RL-based methods, and other ML-based methods. Here, we first introduce some interesting techniques of graph embedding related to the IM problem. Then, we mainly introduce the standard process of RL algorithm that takes the current embedding of nodes or networks as state of the RL agent in order to choose next seed node as action. Finally, we review some other latest ML-based methods to address it.
    
\subsection{Graph Embeddings}
    The GNNs introduced in Section \uppercase\expandafter{\romannumeral5} are very suitable for embedding social networks, which exhibit considerable power in many applications. Also, plenty of variants are developed for diverse purposes. We describe a few of them which are applicable to the IM problem. Aiming at the problem of ignoring hierarchical structure in graph embedding, Ying \textit{et al.} \cite{ying2018hierarchical} propose a deep GNN model named \textit{DiffPool}, which maps the whole graph to a single embedding vector in an end-to-end differentiable manner with hierarchical structure encoded. \textit{DiffPool} has an architecture of multiple layers of GCNs, In each layer, the nodes of the input network are first encoded into embedding vectors, and then clustered according to their embedding vectors to form a coarser network, in which each cluster is regarded as a new node, and edges between clusters are preserved. The input network is coarsened iteratively to a single embedding vector of the entire network, and the hierarchical features are encoded. The model is first raised for graph classification, and achieves $5 \mbox{-}10$\% improvement in accuracy compared with the SOTA benchmarks. Then, Kamarthi \textit{et al.} \cite{kamarthi2020influence} adopt it to solve the IM problem under an unknown graph. Feng \textit{et al.} \cite{feng2018inf2vec} propose a social influence embedding method against sparsity of observed propagation data. Under the tasks of node activation and influence diffusion prediction, it outperforms the SOTA baseline algorithms, and it is applicable to the IM problem with unknown diffusion probabilities \cite{panagopoulos2020influence}. They try to get the representation of the node that encoding influence context, and then compute the similarity of every two node embeddings as an indicator of edge propagation probability between them, instead of directly estimating the edge influence probability, which is affected by the sparse edge propagation data. In order to encode the influence context, they devise \textit{Inf2vec} embedding algorithm, which utilizes the \textit{word2vec} model \cite{mikolov2013efficient} \cite{mikolov2013distributed} to learn node embedding from the network structure, observed influence diffusion history, and similarity of user interest. Specifically, the network structure and influence diffusion can be extracted from the results of a random walk strategy, and the similarity information of user interest is characterized by sets of users who perform the same action. Wilder \textit{et al.} \cite{wilder2019end} design an end-to-end decision-focused learning framework, which is suitable for the problem of selecting an optimal seed set of $k$ nodes such as the IM problem. To meet the need for such combination of learning and optimization problem, this framework does not choose either a separate two-stage way or conventional end-to-end way. Instead, it integrates a differentiable solver for a simpler problem of the true complex problem as a layer of the architecture, and learns the mapping representation from the true problem to the simpler problem for the solver. This method focuses on learning representation rather than optimization, and shows better performance than two-stage or conventional end-to-end methods on a subset selection task. 
    
    In real applications of the IM problem, different scenarios focus on different context information of social networks, such as topic-aware, location-aware, etc, which gives new challenges to embedding algorithms. Here we briefly introduce some context-aware embedding methods. Wu \textit{et al.} \cite{wu2019neural} propose an embedding approach, \textit{DiffNet}, modeled by a layer-wise influence diffusion structure, taking the evolving nature of social networks into account. As a graph based model, \textit{DiffNet} shows better performance than the non-graph based recommendation models. Wu \textit{et al.} \cite{wu2020diffnet++} then improve \textit{DiffNet} structure and propose a \textit{DiffNet++} model, which considers the social network and interest network of users to learn the user embedding in recommendation systems. It embeds both user-item graph and user-user graph for social recommendation, and is shown to be effective by the experiment results. Khajehnejad \textit{et al.} \cite{khajehnejad2020adversarial} propose an adversarial embedding method, in which an auto-encoder is used for node embedding and a discriminator is used for discriminating sensitive attributes, so as to achieve similar distributions among sensitive attributes, thus protect sensitive attributes in fair IM scenarios. Nishad \textit{et al.} \cite{nishad2020graphreach} devise a position-aware GNN called \textit{GraphReach} to improve embedding effectiveness by considering positional information. Yu \textit{et al.} \cite{yu2021influence} devise an influence-aware GNN which can encode the influence of a node in a network, facilitating subsequent tasks such as the IM problem. Chen \textit{et al.} \cite{chen2022graph} present a comprehensive review of recent works that leverages Graph Representation Learning for popularity prediction problem, in which a lot of ML-based techniques are used to study information diffusion. It can be applied to the IM problem as well.
     
     To sum up, graph embedding techniques in the IM problem can learn IM related features, such as hierarchical features, community features and various context related information. For a specific application scenario, the existing embedding models can be tailored or advanced according to the specific objective or trait to adapt to the problem need.
    
\subsection{RL-Based IM Algorithms}
    \subsubsection{Single-Agent RL without Pruning} Some methods attempt fusing DRL into greedy the IM framework, estimating the marginal influence gain of each node by deep neural network and selecting the most influential one each time. Li \textit{et al.} \cite{li2019disco} extend the method in \cite{khalil2017learning} for generic IM problem and propose the DISCO framework composed of GNNs and DQN, where the action is designed as selecting a node into seed set and the reward is its marginal influence gain. In contrary to hand-crafted feature, DISCO embeds the nodes in the graph network as feature vectors through a recurrent GNN named \textit{structure2vec} \cite{dai2016discriminative} which takes account of the connection information of nodes in the graph as well as the partial solution seed set $S$ resulting in better generalization ability. Once the seed set $S$ updated, DISCO would update the embedding vectors $\mu_{v}^{(i+1)}$ as follows:
    \begin{equation}\label{eq17}
        \mu_{v}^{(i+1)}=\text{ReLU}\left(\alpha_1 \sum\nolimits_{u\in \mathcal{N}(v)} \mu_{u}^{(i)}+ \alpha_2 \sum\nolimits_{u\in \mathcal{N}(v)}\text{ReLU}\left(\alpha_3 w(v,u)\right)+\alpha_4 a_v\right),
    \end{equation}
    where ReLU refers to the Rectified Linear Unit of a neural network, $\mu_{v}^{(i)}$ is the embedding vector of node $v$ during the $i^{\text{th}}$ iteration, $\mathcal{N}(v)$ is the set of neighbors of node $v$, $w(v,u)$ is the weight of edge $(v,u)$ which can correspond to diffusion possibility under the IC model, $a_v$ is a scalar of 1 or 0 to indicate whether node $v$ is in partial seed set $S$ currently, and $\alpha_1$, $\alpha_2$, $\alpha_3$, and $\alpha_4$ are trainable parameters which can be learned from the given IM problem. DISCO defines a Q-function of DQN as follows, which is the approximate function mapping from the embedding vectors of $v$ and all other nodes to the marginal influence of node $v$ given seed set $S$. That can be defined as follows:
    \begin{equation}\label{eq18}
        Q(v,S,\Theta)=\beta_{1}^{T}\text{ReLU}\left(\left[\beta_2 \sum\nolimits_{u\in V}\mu_{u}^{(I)},\beta_3\mu_{v}^{(I)}\right]\right),
    \end{equation}
    where $[\cdot,\cdot]$ denotes the concatenation operator, $\beta_1$,$\beta_2$, and $\beta_3$ are learnable parameters, and $\Theta$ represents the set of all related learnable parameters, i.e. $\alpha_1,\cdots,\alpha_4$ and $\beta_1,\cdots,\beta_3$. As we known, computing the marginal influence is \#P-hard. DISCO trains a Q-function to predict the marginal influence gain of a node based on the generalized node representation, avoiding sampling the stochastic diffusion process of traditional approaches which is costly. For application of the trained model, the authors propose a seed selection method that would selects all $k$ nodes for a seed set at once according to decreasing Q-values given empty partial solution set instead of picking only one node at a time by hill-climbing greedy strategy, which accelerates the process. They claim that the order of the nodes with respect to their Q-values remain almost unchanged whether or not they do hill-climbing greedy strategy with multiple-round computing of network embedding and Q-values. DISCO outperforms traditional methods like IMM \cite{tang2015influence} and SSA \cite{nguyen2016stop} in terms of computational efficiency with comparable influence quality of the seed set. Even though this work basically follows the procedures of S2V-DQN \cite{khalil2017learning} and its experiments are not robust and convincing, it is an important work to address the IM problem by using the latest RL algorithm.
    
    Soon afterwards, Li \textit{et al.} \cite{li2022piano} (same as the authors of DISCO) proposed the PIANO framework, which is an evolved version of DISCO. Its general idea is the same as DISCO's, but it perfects the algorithm design and theoretical analysis, and supplements a large number of comparative experiments. Based on the framework of DISCO, Wang \textit{et al.} \cite{wang2021reinforcement} propose the IMGER for large-scale and dynamic social networks. The IMGER utilizes GAT \cite{velivckovic2018graph} for network embedding, which can learn the weights of edges from data by attention mechanism instead of pre-defined weights related to the diffusion model in order to fit diverse realistic social networks. Double DQN (DDQN) \cite{van2016deep} algorithm in DRL is used to learn a greedy policy and select the seed set to maximize social influence in a similar manner to DISCO. The results show that the IMGER demonstrates superior performance in both the social diffusion prediction and the quality of IM solution set. To enhance the generalization ability, Chen \textit{et al.} \cite{chen2022touplegdd} propose an end-to-end framework called Three Coupled GNNs and Double DQN (ToupleGDD) to solve the IM problem by incorporating \textit{DeepWalk} and three coupled GNNs for network embedding and DDQN for parameters learning. They claim that ToupleGDD can be trained on small randomly generated graphs and tested on totally and completely diversified networks Their results have been validated in different real world datasets, whose performance is very close to SOTA IM algorithms. Actually, this is a milestone work of pursuing a stronger generalization ability by applying DRL in the IM problem.
        
    \subsubsection{Single-Agent RL with Pruning} To enhance the efficiency of DRL-based IM algorithms, some methods integrate network pruning techniques to navigate the computing resources focusing on more important nodes or subnetworks. Shown as Section \uppercase\expandafter{\romannumeral5}, Manchanda \textit{et al.} \cite{manchanda2020gcomb} propose the GCOMB framework that can be used to solve the IM problem as well through introducing a node quality evaluation mechanism. Unlike the end-to-end architecture in DISCO, GCOMB has two separate components of network embedding and Q-learning using both supervised learning and RL to train the model. One component is a convolution-based GNN named GraphSAGE \cite{hamilton2017inductive} employed to prune poor nodes and learn embedding of good nodes in a supervised manner, where the quality of a node $v$ takes into account the influence overlapping between partial solution $S$ and $v$. The other component is the Q-learning RL architecture which only considers the good nodes for solution. They conduct a large number of comparative experiments with the SOTA IM algorithms, including S2V-DQN \cite{khalil2017learning}, GCN-TREESEARCH \cite{li2018combinatorial}, Greedy (Algorithm \ref{a1}), CELF \cite{leskovec2007cost}, IMM \cite{tang2015influence}, and OPIM-C \cite{tang2018online}. It shows that the pruning method largely reduces the computational cost, making it scalable to billion-sized graphs with exponentially less time consumption than other algorithms. The GCOMB can be generalized to unseen graphs of significantly larger sizes and convincingly outperform the SOTA IM algorithms. Ireland \textit{et al.} \cite{ireland2022lense} propose a framework called LeNSE to prune graph in order to further reduce the running time for CO problems. Specifically, it identifies qualified subgraphs from the original entire graph, which can represent the original one in terms of implicating the solution, and adopts heuristics in those subgraphs to get a solution within reasonable time. The results show that the LeNSE outperforms the corresponding pruning methods in GCOMB.
            
    \subsubsection{Multi-Agent RL} There are also some efforts to model the IM problem as group dynamics of multiple learning agents. Khomami \textit{et al.} \cite{daliri2018minimum} introduce a learning automata based algorithm for solving the minimum positive influence dominating set problem and apply it to address the IM problem. It assigns each node a learning agent to cooperate with each other for deciding the seed set. Singh \textit{et al.} \cite{singh2019lapso} utilize the two-hop influence spread to approximate the actual influence spread, and adopt learning automata based method to maximize influence. Liu \textit{et al.} \cite{liu2021multiple} propose a MAIM algorithm which uses multi-agent Memory Separated DQN to improve training efficiency without network embedding techniques. It designs the action as either accepting current node and adding it into seed set or dropping it without reconsideration by explicitly considering the influence overlapping. Ma \textit{et al.} \cite{ma2022influence} combines the RL with evolutionary algorithms, using it to deal with a bunch of DQNs which make decisions on the selection of next seed node. By using evolutionary algorithms, it can effectively compress the searching space of training in reinforcement learning, thus achieving the purpose of improving efficiency.

\subsection{Other ML-Based IM Algorithms}
    There are some studies using other ML-based methods (not RL-based methods), including using node or network embeddings as well as supervised or unsupervised deep learning methods to find influential seed nodes. 
    
    \subsubsection{Unsupervised or Self-Supervised Manner} The complexity of graphs makes it difficult to obtain labeled training data in the IM problem. Thus the majority of ML-based IM algorithms resort to unsupervised or self-supervised learning approaches. Ni \textit{et al.} \cite{ni2021fastcover} reduce the IM problem to a budget-constrained multi-hop dominating set problem. The idea is to learn a scoring mechanism for all nodes such that each node's score can represent its topology-integrated influence in order to directly choose the top-$k$ nodes as seeds. Specifically, they propose an unsupervised framework, FastCover, constituted of a multi-layer GNN called graph reversed attention network (GRAT) to generate influence score of each node to range $[0,1]$. The parameters of GRAT are trained via optimizing a differentiable loss function in an unsupervised manner. The authors claim FastCover achieves a near-optimal performance with quasi-linear time complexity. To balance the scalablity and high-performance in the IM problem, Zhang \textit{et al.} \cite{zhang2022network} modify a GCN with adapted number of layers to different scales of networks and train it with a self-labeling mechanism by binning nodes regarding degree. Then, they use a distance-based heuristic IM method, taking as input the generated node embeddings, to compute the seed set and achieve better performance for seed set identification tasks with high efficiency than the baselines such as work \cite{keikha2020influence}. Li \textit{et al.} \cite{li2022mahe} propose a deep learning algorithm MAHE-IM for the IM problem based on heterogeneous networks, designing a embedding method to extract high-order representations from trajectories by metapath-based random walk method \textit{metapath2vec} \cite{dong2017metapath2vec}. By comparing two nodes' cosine similarity and setting a similarity threshold for relevance, they figure out a set of relevant nodes corresponding to each node and select top-$k$ node with highest occurrence frequencies among those sets. Extensive experiments are conducted against eighteen SOTA baseline algorithms, showing the effectiveness of MAHE-IM on heterogeneous networks.
    
    The diffusion models with random or uniform influence parameters may suffer from inaccurate estimation of realistic influence diffusion \cite{aral2018social}, which promotes estimating edge probabilities from history data of diffusion cascades. Also, due to the number of parameters to estimate in this case being equal to the large number of edges, overfitting tends to happen. Bourigault \textit{et al.} propose a practical method that forms an edge probability dependently with the pair of embeddings corresponding to the connected influential and susceptible nodes \cite{bourigault2016representation}. Following this idea, Panagopoulos \textit{et al.} \cite{panagopoulos2020influence} propose a model-independent method called CELFIE, introducing a node embedding method to form the activation probabilities between nodes which is used to reconstruct a simpler bipartite network for seed set selection in the IM problem. The embedding model \textit{L-Inf2vec} is proposed based on \textit{Inf2vec} \cite{feng2018inf2vec} which is introduced previously for influencer and susceptible node embeddings, where the influencers refer to nodes sending influence and the susceptible nodes refer to those receiving influence. Then the two parts of a bipartite network are built up by connecting influencers to susceptible nodes where the weight of an edge is the diffusion probability computed with influencer and susceptible node embeddings. Then the IM problem could be solved under this bipartite network by simple neighborhood sampling, with the simulation on traditional diffusion model discarded, which accelerates the computation of influence spread. 
    
    Nevertheless, CELFIE has drawbacks of relying on sampling-based influence spread and not considering conditional probabilities regarding partial seed set, which causes weak spread increment as the size of seed set increases during seed selection. To overcome this, they subsequently propose IMINFECTOR, an extension of CELFIE based solely on diffusion cascades \cite{panagopoulos2020multi}. IMINFECTOR also utilizes the influence and susceptibility node embeddings learned from cascades to form the diffusion probabilities between nodes by a multi-task neural network \cite{caruana1997multitask}, but an expected influence spread is assigned to each node by taking into account the partial seed set. The experiment results show that IMINFECTOR outperforms several competitive methods including CELFIE in spread quality or time efficiency. Still, due to the chronological order in which IMINFECTOR retrieves the context of nodes, there exists the possibility that less influential nodes are actually selected as candidate nodes. To overcome this limitation, Liu \textit{et al.} \cite{liu2022influence} propose LDRLIM that obtains the influence of the node according to node embeddings. LDRLIM uses Discount-degree Descending (DED) search strategy to generate the candidate nodes set and IC model based random Walk from candidate nodes to produce node propagation sequences for learning node embeddings. The IC model is taken and the diffusion probability of an edge depends on the $2$-norm similarity between influencer and susceptible node embeddings, which is followed by CELF method for the IM solution. Xie \textit{et al.} \cite{xie2022hierarchical} also assume unobservable diffusion probabilities under the IC model, and propose an approach to predicting probabilities by neural network and maximizing influence. They first compute the embeddings of nodes by using an MLP model to take as input hierarchical community clusters, node attributes and general network structure. The diffusion probability of an edge can then be calculated based on the $2$-norm similarity between connected nodes. With the diffusion probabilities, the seed set is figured out via greedy algorithm and outperforms the SOTA methods for the IM problem under attributed networks.
        
    \subsubsection{Supervised or Semi-Supervised Manner} Despite the challenges of collecting labeled data for the IM problem, some studies make an effort to construct semi-supervised or even supervised deep learning forms for performance concern. Shen \textit{et al.} \cite{shen2017leveraging} propose a Cross-Network Graph Sparsification (CNGS) algorithm to prune edges of large-scale graphs for accelerating downstream IM algorithms. CNGS is trained in a semi-supervised on a source network and exploits the knowledge learned to guide the removal of edges in a target network with minor impact on influence diffusion. By removing edges unlikely to be influential, the existing IM algorithms can be implemented quite faster due to a sparser network than before, while performance can barely suffer degradation. The results demonstrate at least $39$\% time saved with only less than $10$\% performance reduced on several real-world public datasets. Soon after this work, they further improve the method and introduce fuzzy self-training approach to enhance the cross-network prediction performance \cite{shen2019cross}. Kumar \textit{et al.} \cite{kumar2023influence} convert the IM problem into a pseudo deep learning regression problem that trains a generalizable model to transfer to unseen networks. With a graph-based LSTM, they extract the centrality-based feature vectors of nodes from a training network as input, and compute the influence spread of nodes under the susceptible-infected-recovered (SIR) and IC model as target labels, then transfer the trained LSTM to a new target network to predict the influence of each node for top-$k$ selection of seed set. Similarly, Kumar \textit{et al.} \cite{kumar2022influence}, the same team as the previous, propose another supervised IM algorithm, still converting to a regression problem but adopting the node embedding algorithm \textit{struc2vec} \cite{ribeiro2017struc2vec} to extract node representation, and then feed the embedding vectors into a GNN to learn node's influence, outperforming some of classical and latest IM algorithms.
    
    Generally speaking, the design of IM algorithms is much more flexible than the traditional algorithms, involving the integration of deep learning, reinforcement learning, graph representation learning, evolutionary algorithms, combinatorial optimization and other technologies. Making full use of the advantages of each technology, it gradually improves the quality of solutions, reduces time complexity, enhances generalization ability, and lands in real business scenarios. Through the research of IM algorithms and social applications, it also promotes the development of machine learning and combinatorial optimization.
    
\begin{table*}[!t]
        \caption{Summary of Deep Algorithms for the IM and its variant problems \\(CLT, CIC and MCTS are short for Competitive Linear Threshold, Competitive Independent Cascade respectively, and Monte-Carlo Tree Search.)}
        \label{tab1}
        \resizebox{\textwidth}{90mm}{
        \begin{tabular}{|l|l|p{9em}|p{5em}|p{10.5em}|l|}
        \hline
        Year & Author & Problem & Diffusion model & Embedding & Training method \\
        \hline
        2017 & Shen \textit{et al.} \cite{shen2017leveraging} & \multirow{22}{*}{Vanilla IM} & IC & N.A. & semi-supervised \\
        2018 & Khomami \textit{et al.} \cite{daliri2018minimum} &  & LT & N.A. & learning automata \\
        2019 & Shen \textit{et al.} \cite{shen2019cross} &  & IC & N.A. & semi-supervised \\
        2019 & Li \textit{et al.} \cite{li2019disco} &  & IC & \textit{structure2vec} & DQN \\
        2019 & Liu \textit{et al.} \cite{liu2019predicting} &  & IC/LT & GCN & supervised \\
        2019 & Singh \textit{et al.} \cite{singh2019lapso} &  & IC/LT & N.A. & learning automata \\
        2020 & Manchanda \textit{et al.} \cite{manchanda2020gcomb} &  & N.A. & GraphSAGE & Q-learning \\
        2020 & Panagopoulos \textit{et al.} \cite{panagopoulos2020influence} &  & N.A. & \textit{Inf2vec} & self-supervised \\
        2020 & Panagopoulos \textit{et al.} \cite{panagopoulos2020multi} &  & N.A. & Multi-task NN & self-supervised \\
        2021 & Wang \textit{et al.} \cite{wang2021reinforcement} &  & N.A. & GAT & DDQN \\
        2021 & Liu \textit{et al.} \cite{liu2021multiple} &  & IC & N.A. & Multi-agent DQN \\
        2021 & Ni \textit{et al.} \cite{ni2021fastcover} &  & N.A. & GAT & unsupervised \\
        2022 & Zhang \textit{et al.} \cite{zhang2022network} &  & IC & GCN & self-supervised \\
        2022 & Li \textit{et al.} \cite{li2022mahe} &  & IC/LT & \textit{metapath2vec} & unsupervised \\
        2022 & Li \textit{et al.} \cite{li2022piano} &  & IC & \textit{structure2vec} & DQN \\
        2022 & Kumar \textit{et al.} \cite{kumar2023influence} &  & SIR & hand-crafted features & supervised \\
        2022 & Kumar \textit{et al.} \cite{kumar2022influence} &  & IC & \textit{struc2vec} & supervised \\
        2022 & Liu \textit{et al.} \cite{liu2022influence} &  & IC & Multi-task NN & self-supervised \\
        2022 & Xie \textit{et al.} \cite{xie2022hierarchical} &  & IC & MLP & unsupervised \\
        2022 & Ma \textit{et al.} \cite{ma2022influence} &  & IC & LINE & DQN \\
        2022 & Ireland \textit{et al.} \cite{ireland2022lense} &  & IC & GNN & Q-learning \\
        2022 & Chen \textit{et al.} \cite{chen2022touplegdd} & & IC & GNN & DDQN \\
        \hline
        2021 & Ou \textit{et al.} \cite{ou2021active} & \multirow{3}{*}{Blocking IM} & SIS & GCN & DQN \\
        2022 & He \textit{et al.} \cite{he2022graph} &  & LT variant & hand-crafted features & semi-supervised \\
        2022 & Yilmaz \textit{et al.} \cite{yilmaz2022misinformation} &  & SIS & N.A. & multi-agent DDPG \\
        \hline
        2015 & Lin \textit{et al.} \cite{lin2015learning} & \multirow{7}{*}{CIM} & CLT & hand-crafted features & Q-learning \\
        2018 & Yang \textit{et al.} \cite{yang2018dna} &  & CLT & hand-crafted features & MCTS \\
        2018 & Ali \textit{et al.} \cite{ali2018boosting} &  & CLT variant & hand-crafted features & Q-learning \\
        2020 & Ali \textit{et al.} \cite{ali2020addressing} &  & CLT variant & hand-crafted features & DQN \\
        2021 & He \textit{et al.} \cite{he2021reinforcement} &  & CIC & hand-crafted features & Q-learning \\
        2021 & Ali \textit{et al.} \cite{ali2021nedrl} &  & CLT variant & \textit{node2Vec} & DQN \\
        2022 & Ali \textit{et al.} \cite{ali2022leveraging} &  & CLT & hand-crafted features & Q-learning \\
        \hline
        2018 & Yadav \textit{et al.} \cite{yadav2018please} & \multirow{2}{*}{Conting.-aware IM} & N.A. & N.A. & POMDP planner \\
        2021 & Chen \textit{et al.} \cite{chen2021contingency} &  & IC & \textit{structure2vec} & DQN \\
        \hline
        2021 & Meirom \textit{et al.} \cite{meirom2021controlling} & \multirow{3}{*}{Dynamic IM} & IC/LT & GNN & PPO \\
        2022 & Mendon{\c{c}}a \textit{et al.} \cite{mendoncca2022efficient} &  & IC variant & \textit{structure2vec} & DQN \\
        2022 & Sheng \textit{et al.} \cite{sheng2022dynamic} &  & IC/LT & \textit{DynamicTriad} & semi-supervised \\
        \hline
        2019 & Li \textit{et al.} \cite{li2019deep} & \multirow{3}{*}{Location-aware IM} & N.A. & hand-crafted features & supervised \\
        2020 & Zhang \textit{et al.} \cite{zhang2020geodemographic} &  & self-defined & hand-crafted features & unsupervised \\
        2022 & Jin \textit{et al.} \cite{jin2022im2vec} &  & IC & \textit{All2Vec} \cite{jin2022im2vec} & unsupervised \\
        \hline
        2020 & Keikha \textit{et al.} \cite{keikha2020influence} & Multilayer IM & IC/LT & CARE \cite{keikha2018community} & unsupervised \\
        \hline
        2020 & Kamarthi \textit{et al.} \cite{kamarthi2020influence} & \multirow{4}{*}{Graph-unknown IM} & IC & \textit{DiffPool}+\textit{DeepWalk} & DQN \\
        2021 & Tran \textit{et al.} \cite{tran2021meta} &  & IC/WC & Siamese NN \cite{bromley1993signature} & unsupervised \\
        2021 & Li \textit{et al.} \cite{li2021claim} &  & IC & \textit{DiffPool}+\textit{DeepWalk} & DQN \\
        2022 & Wu \textit{et al.} \cite{wu2022gac} &  & IC & GraphSAGE+\textit{DiffPool} & DDPG \\
        \hline
        2019 & Tian \textit{et al.} \cite{tian2019learning} & \multirow{4}{*}{Topic-aware IM} & IC & \textit{Diffusion2Vecv} & DDQN\\
        2020 & Tian \textit{et al.} \cite{tian2020deep} &  & IC/LT & \textit{Diffusion2Vecv} \cite{tian2020deep} & DDQN\\
        2020 & Huang \textit{et al.} \cite{huang2020recurrent} &  & CIC variant & N.A. & supervised \\
        2021 & Huang \textit{et al.} \cite{huang2021competitive} &  & CIC variant & N.A. & supervised \\
        \hline
        \end{tabular}
        }
    \end{table*}
    
\section{ML-Based Variants of Influence Maximization}
    The variant IM problems combine the vanilla IM problem with diverse traits of practical scenarios and they are of research value for real-life applications. In this section, we survey the latest applications for the variant IM problems, and focus more on those works adopting ML-based methods. In fact, these variant IM problems have been defined and studied in the era of traditional algorithms. If you want to know more about their definitions and are interested in traditional algorithms for solving variant IM problems, please check Li's survey \cite{li2018influence}. It is very challenging to summarize this part of the work, which mainly involves two difficulties. First, how to classify these variant IM problems; secondly, how to accurately judge whether it belongs to the traditional algorithm or the ML-based algorithm, and summarize its adopted technologies. The taxology and its corresponding representatives are summarized in Table \ref{tab1}.

\subsection{Competitive Influence Maximization (CIM)}
    The competitive IM problem assumes multiple parties in a social network compete against each other in propagation of information, e.g. one mobile phone provider would make a suitable marketing policy to maximize the number of its customers when there are other providers giving policies to compete in the same social network. The competitive IM (CIM) problem can be divided into two cases: Known opponent strategies and Unknown opponent strategies. Bharathi \textit{et al.} \cite{bharathi2007competitive} first define the competitive IM problem as a game problem. Thus, the research in traditional algorithms for this area mainly focused on algorithmic game theory and submodular optimization. Recently, there are a lot of ML-based methods used to address it as follows.
    
    Lin \textit{et al.} \cite{lin2015learning} propose solutions for both two competitive IM cases of known and unknown opponent strategy. If known, they adopts Q-learning techniques of RL considering multi-round situation, training an agent to learn a strategy to defeat the opponent's strategy, or else a game theoretical solution is used to seek the Nash equilibrium as the best strategy which a rational agent should act on. To avoid extra computational overhead caused by high dimensional state and action space of RL, they design hand-crafted features as the state representation to represent the current state of social network and borrow the idea of meta-learning to choose a candidate strategy rather than a node as an action. The rewards are designed as the relative marginal influence gain compared to the opponents, i.e. the difference of activated nodes between parties in the last round. Since RL agents take a large amount of random exploration actions to find the best policy, which is time consuming, Ali \textit{et al.} \cite{ali2018boosting} develop based on Lin's work for the Time-Constrained CIM (TC-CIM) scenario where the objective of each party in TC-CIM scenario is to design a strategy that can maximize its overall influence spread within its very limited time budget. They use similar framework with hand-crafted network features as well as Q-learning under Time-Varing Competitive LT diffusion model and apply transfer learning with the help of other knowledge to reduce the training time. Yang \textit{et al.} \cite{yang2018dna} extend the work in \cite{lin2015learning} to a general situation that the parties could select nodes and propagate by turns, and propose an RL-based framework to solve the multi-round multi-party IM problem. With the similar idea in \cite{ali2018boosting}, Ali \textit{et al.} \cite{ali2020addressing} model the CIM problem under topology-unknown network scenario as an RL problem, then apply DQN to learn when to explore the unknown network and how to select seed set, achieving better performance than heuristic methods. Next, He \textit{et al.} \cite{he2021reinforcement} study competitive opinion maximization in signed social network where negative weights of edges are introduced to model the relationship of dislike or distrust. They exploit Q-learning to select nodes against unknown opponent strategy. Ali \textit{et al.} \cite{ali2021nedrl} propose a DRL-EMB framework through adopting network embedding method \textit{node2vec} \cite{grover2016node2vec} and DQN algorithm to formulate evolving social networks instead of static network. Ali \textit{et al.} \cite{ali2022leveraging} also improves the work in \cite{lin2015learning} by integrating transfer learning in RL method on the competitive LT model. To facilitate the transfer of model from source to target network, they normalize the two networks and propose a transfer learning method to accommodate the transferred agent in regard to heterogeneity between networks and agent settings. Better or similar performance is claimed for this transfer learning based model than the baseline model trained from scratch, with greatly less time consumption.
        
\subsection{Topic-Aware Influence Maximization}
    The topic-aware IM problem takes into account people's interests in topics when increasing the spread of certain topic-related information. It uses topic to represent item characteristics and users' interests, formulated to find the optimal seed set that maximizes the specifically defined topic-aware influence. The research in traditional algorithms in this area can be categorized into the IM for topic relevant targets \cite{li2015real} \cite{nguyen2016cost} and for topic-dependent diffusion \cite{chen2015online} \cite{aslay2014online}. The first one aims at maximizing the influence over users related to a specific topic, and the second one is to adjust diffusion models such that the diffusion probability is topic-dependent. Recently, there are a lot of ML-based methods used to address it as follows.
    
    Tian \textit{et al.} \cite{tian2019learning} apply a similar method of DISCO \cite{li2019disco} to the topic-aware IM problem where people have different probabilities to be activated in different topics. They also extend the work in \cite{khalil2017learning}, using network embedding and DDQN framework. The embedding model called \textit{Diffusion2Vec} exploits the basic architecture of \textit{structure2vec} \cite{dai2016discriminative}, but modified it by incorporating a weighted vector, such as $\{<music,0.7>,<sport,0.3>,<politics,0.0>\}$ to represent the extent of interest in specific topics of a node, into the feature attribute vector $x_v$ in Eqn. (\ref{s2v}). In addition, they adopt a data-driven method to quantify the propagation parameters of diffusion model rather than assigning random values from range $[0,1]$ in order to approximate the real-life situations. The IC model is adopted, and the diffusion parameters of edges can by calculated as follows:
    \begin{equation}
        p_{uv}^{\tau} = \left(\gamma_1 w_{uv}+\gamma_2 sim(u,v)+\gamma_3 B_{v}^{\tau} \right)/3,
    \end{equation}
    where $p_{uv}^{\tau}$ denotes the propagation probability from $u$ to $v$ in terms of topic $\tau$, $w_{uv}$ denotes the initial weights of edge from $u$ to $v$, $sim(\cdot,\cdot)$ is the cosine similarity function between the representation of two nodes, $B_{v}^{\tau}$ represents the weight of interest towards topic $\tau$, and $\gamma_1,\gamma_2,\gamma_3\in (0,1)$. 
    Tian \textit{et al.} \cite{tian2020deep} (the same authors of \cite{tian2019learning}) subsequently augment the theoretical analysis and extend the experiments to the problem under the LT model. The performance of this RL-based algorithm is shown superior to those manually designed topic-aware IM baselines.
    
    Huang \textit{et al.} \cite{huang2020recurrent} consider the diffusion spread of a certain entity where its competitive or complementary products have already been promoted across the same network, which can be regarded as a topic-aware variant where the activation probability of a complementary topic should be larger than that of a competitive topic. This is called a follower-based IM problem by the authors. To solve this, they propose an algorithm that recurrent neural variational model is adopted to update the evolving thresholds in LT model by learning the dynamics of multiple-entity interactions and cascades, then seed set can be computed via a greedy algorithm. On the basis of this work, Huang \textit{et al.} \cite{huang2021competitive} soon afterwards propose a similar method as \cite{huang2020recurrent} to solve the follower-based
    IM problem, but consider adjusted IC model and enhance the influence probability inference model, using modified LSTM model in place of recurrent neural variational model. Experiments show their method outperforms the SOTA methods in terms of prediction accuracy and efficiency.

\subsection{Contingency-Aware Influence Maximization}
    Yadav \textit{et al.} \cite{yadav2018please} first propose the contingency-aware IM problem where it is uncertain whether a node is willing to accept the invitation when invited to become a seed node. This was inspired by spreading the awareness of HIV prevention among homeless youth. In the past, this problem has been addressed by applying Partially Observable Markov Decision Process (MDP) \cite{yadav2016using} and Greedy Algorithm \cite{wilder2018end}. However, the solution to the contingency-aware IM problem needs to be recomputed when the underlying network changes, and the stakeholders have no high-performance computing resource. In other words, it requires higher generalization ability of the algorithm.
    
    Thus, Chen \textit{et al.} \cite{chen2021contingency} re-study the contingency-aware IM problem by using DRL technique. To overcome the difficulties of how state incorporate the uncertainty information and how reward can be computed because of the uncertainty in node status, they proposed a new MDP formulation, state-abstraction component, and reward-shaping component, which constitute the RL4IM framework through network embedding model \textit{structure2vec} and DQN. Besides, the authors assume historical data of the users' choices are available. The historical data can provide some prior knowledge of the willingness of nodes that can be used to build up the states and actions of DQN. The RL4IM can conduct a seed selection strategy on a laptop within seconds after sufficient training.
        
\subsection{Graph-Unknown Influence Maximization}
    The fact that the social network connection is sometimes unknown triggers the IM problem in graph-unknown social networks, of which the algorithms normally integrate social network discovery methods. For example in public health, the network dataset is typically gathered by in-person intervention, which is not feasible for use to solve the IM problem. Thus, in some scenarios, obtaining the entire social graph is costly because it has to survey multiple people. To deal with it, Wilder \textit{et al.} \cite{wilder2018end} propose a CHANGE algorithm that is based on friendship paradox by surveying a random node and one of its neighbors. The subgraph obtained by surveying a small number of nodes used to select a seed set for the IM problem. Subsequently, Kamarthi \textit{et al.} \cite{kamarthi2020influence} formulate the same problem of network discovery by MDP, and developed an extension to DQN architecture named Geometric-DQN to learn policies for network discovery by extracting relevant graph properties. It can learn global state representation by using \textit{DiffPool} \cite{ying2018hierarchical} for the discovered graph and local action representation by using \textit{DeepWalk} \cite{perozzi2014deepwalk} for node embedding. They show that their RL-based algorithm has $7$-$13$\% improvement over the SOTA sampling algorithm, CHANGE.
    
    However, Kamarthi's RL algorithm has to perform multiple interactions with the environment, which is very costly in real world applications. To increase the sample efficiency, Li \textit{et al.} \cite{li2021claim} formulate the network discovery and IM problem as a goal directed learning problem \cite{andrychowicz2017hindsight} \cite{nair2018visual} with unknown goal state. Then, they propose a Curriculum Learning Policy for Influence Maximization (CLAIM) by using curriculum guided Hindsight Experience Replay \cite{fang2019curriculum} and goal directed Geometric-DQN to learn efficient policies, where the influence spread can be improved more $7.51$\% than existing methods. To further study this problem, Tran \textit{et al.} \cite{tran2021meta} introduce an end-to-end framework, IM-META, to deal with the IM problem in attributed networks with unknown topology, where they formulate it as an optimization problem to select both seed set and query set by three steps: network inference, reinforced weighted graph generation, and query node selection. They show that through query only $5$\% of nodes, it is enough to get an effective seed set comparing with other benchmark IM algorithms. Wu \textit{et al.} \cite{wu2022gac} consider the influence strength between users is unknown and proposed an end-to-end RL-based framework, Geometric Actor-Critic (GAC) to effectively find policies for incentive allocation, a variant IM problem, under limited budgets. They use GraphSAGE and \textit{DiffPool} models to learn the representation of nodes and entire graph, then feed the representation into Deep Deterministic Policy Gradient (DDPG) \cite{lillicrap2015continuous} to compute the action of which node to choose for the seed set. Even though the GAC outperforms other baselines, it needs two node matrices as input, which is too large to handle extremely large social networks.
        
\subsection{Dynamic Influence Maximization}
    In real applications, the social network is dynamic and keeps evolving. For example, new friendship forms, where a number of edges are added or removed at any time. It leads to additional challenges for the IM problem. Ohsaka \textit{et al.} \cite{ohsaka2016dynamic} first propose a dynamic mechanism under the IC model in evolving graphs, which can give a seed set in real time when there are node or edge updating. In addition to adopting traditional sampling-based approaches, there are some ML-based methods emerging recently.

    Meirom \textit{et al.} \cite{meirom2021controlling} consider studying the problem of controlling partially-observed diffusion on an evolving graph, and based on it, present the partially observed IM problem. They formulate it as a partially observed MDP, then selecting a subset of nodes (seed set in the IM problem) for dynamic intervention as a ranking problem. In order to deal with underlying dynamics, they design an actor-critic RL algorithm which contains two GNN modules, one updates the node representation based on dynamics, and the other cares about long range information diffusion. In addition to the IM problem, this framework can be used in epidemic tests, traffic control, and active sensing. Mendon{\c{c}}a \textit{et al.} \cite{mendoncca2022efficient} design the Spatio-Temporal Influence Maximization (STIM) in evolving networks there where the connectivity between nodes is constantly changing, which is a model trained with RL and graph embedding (\textit{structure2vec}) aiming to find a suitable moment to begin information diffusion from a seed set such that its influence spread can be maximized. They claim that STIM outperforms the greedy agent and can learn complex patterns hidden in evolving networks without supervised data or handcrafted strategies. Sheng \textit{et al.} \cite{sheng2022dynamic} solve the dynamic IM problem by adopting embedding method \textit{DynamicTriad} \cite{zhou2018dynamic} as well as modified seed selecting algorithm in \textit{DeepIM} \cite{keikha2020influence}. Under the dynamic network, the embeddings would take into account the dynamics of edge-add/remove and weight-change with time, then an updating algorithm is designed to update the seed set by incorporating the change of node insertion and deletion. This is a pioneering work of applying ML-based methods to address dynamics in social networks, and it can be changed to those scenarios under different budget constraints \cite{tong2016adaptive} \cite{de2020efficient}.
        
\subsection{Multi-Layer Influence Maximization}
    In addition to spread in a single social network, in reality, information can also propagate across multiple heterogeneous networks via shared users \cite{nguyen2013least} \cite{gaeta2018model} at the same time, which means that users can be influenced by those users from other social networks \cite{zhan2016discover}. This motivates the studies of the multi-layer IM problem. This problem has been studied by use of traditional algorithm, such as \cite{guo2020multi}, and recently studied by used of ML-based methods.
    
    Keikha \textit{et al.} \cite{keikha2020influence} propose a \textit{DeepIM} method which adopts the community-aware method CARE \cite{keikha2018community} to embed the nodes for the IM problem under the interconnected networks scenario. By measuring the similarity of embeddings, the algorithm chooses $r$ most similar nodes for each node as its similarity set, then figures out the top $k$ nodes that are contained by the most sets as the seed set. Besides the improved performance compared with Greedy \cite{kempe2003im}, CELF++ \cite{goyal2011celf++}, SIMPATH \cite{goyal2011simpath}, and LDAG \cite{chen2010scalable_lt}, the experimental results indicate the importance of shared nodes between different networks, around $20$\% of which are selected as the seeds. Katukuri \textit{et al.} \cite{katukuri2022cim} design an algorithm to identify the influential nodes in multi-layer networks by detecting nodes with strong connections in both inter-layers and intra-layers, and also employ node pruning technique to improve the computing efficiency.
        
\subsection{Location-Aware Influence Maximization}
    The location-aware IM problem is defined in location-based social networks, which aims at maximizing the influence spread of location-relevant users rather than any users in vanilla IM problem. Li \textit{et al.} \cite{li2014efficient} is the first to consider the location-aware IM problem under the IC model, selecting a seed set to activate maximum number of users in a geographical region. Then, a series of researches exploit traditional proxy-based \cite{wang2016distance}  or sampling-based \cite{song2016targeted} algorithms to deal with the local-aware IM problem. Recently, there have been several ML-based frameworks to address similar problems defined in geo-networks.
    
    Li \textit{et al.} \cite{li2019deep} study a vehicle selection problem by following the location-aware IM problem, which selects the most suitable candidate vehicles for mobile advertising in order to achieve maximum influence in vehicle networks where spatial and temporal centrality is considered as key factor for information diffusion. They adopt multi-layer perception (MLP) to predict the future centrality and trajectories of vehicles, and thus adopt preferable seed vehicles. Zhang \textit{et al.} \cite{zhang2020geodemographic} defines a geo-demographic IM problem, which is to find a set of locations within a budget that maximize expected reach given a distribution of population and point-to-point movement statistics. They first give an approximation algorithm to address this problem but it is not efficient. Then, they propose an ML-based variant, NN-Sower, that uses randomization and deep learning to improve efficiency with a slight loss of quality. Jin \textit{et al.} \cite{jin2022im2vec} want to incorporate the user preferences and improve the efficiency in location-based IM problem. They propose a new framework, \textit{IM2Vec}, that includes a representation learning model, \textit{All2Vec}, to capture user user preferences and applies the technique of reverse influence sampling to get an efficient approximation algorithm with $(1-1/e-\varepsilon)$ approximation.
        
\subsection{Blocking Influence Maximization (Rumor Blocking)}
    The online social networks have dramatically changed people’s lifestyles, but also have caused widespread misinformation like rumors and fake news. Thus, the blocking IM problem has emerged, which aims to prevent or alleviate the spread of misinformation \cite{zannettou2019web}. The blocking IM problem is usually of two categories according to employed strategies: spreading truth against misinformation (positive cascades) \cite{budak2011limiting} \cite{guo2020multi} and removing key nodes or edges to suppress the propagation of misinformation \cite{fan2013least} \cite{khalil2014scalable}.
    
    \subsubsection{Positive Cascades}
    Instead of known diffusion model, Tong \cite{tong2020stratlearner} considers the situation where misinformation propagates under an unknown Triggering model and proposes a learning method called StratLearner to select $k$ protector nodes given certain attacker nodes to diminish the spread of misinformation. Given a historical data set of the target social network containing attacker-protector set pairs $\{(M,P)\}$, where the protector set $M$ can be regarded as the label corresponding to the attacker set $P$ for supervised learning, however, it cannot cover all the combinations of nodes in a network. Thus, StratLearner defines a scoring function of $M$ and $P$ to estimate the utility, with a parameterized form as $f(M,P)=\vec{w}^{T}\vec{G}(M,P)$ where $M$ denotes a set of attacker nodes, $P$ denotes a optimal or sub-optimal set of protector nodes, $\vec{G}(M,P)\in \mathbb{R}^{l}$ denotes a feature function extracted from $m$ random sampled subgraphs and $\vec{w}\in \mathbb{R}^{l}$ is a vector weights to learn through a structural Support Vector Machine (SVM). The score of $P$ given $M$ is designed to indicate its prevention effect related to the number of nodes that are not influenced by $M$ after cascade, and it can be used to compute the marginal gains of nodes in greedy framework to find out the optimal seed set. StratLearner realizes set function learning with sets as input and output, and outperforms Naive Bayes, MLP, GCN and Deep Set Prediction Networks (DSPN) \cite{zhang2019deep}. With the same scenario of selecting $k$ seeds to conduct a positive cascade diminishing the spread of misinformation, He \textit{et al.} \cite{he2022graph} extend the LT model of information diffusion considering the opinions and confidence of users, and modify a GCN for directed graphs to select the influential nodes for restrain the spread of rumor. Yilmaz \textit{et al.} \cite{yilmaz2022misinformation} tackle the fake news propagation problem from a game-theoretic perspective as well as an multi-agent DRL perspective, the former of which models the nodes in the network as actual players while the latter of which adopts multi-agent DDPG to learn strategy of selecting seed set.
    
    \subsubsection{Removal of Key Nodes or Edges}
    Active screening aims to find potentially infected individuals and provide testing and treatment in order to slow down the spread of disease. It can be regarded as targeting the seed set of people that might maximize the spread of disease, which is also a kind of variant IM problems. Ou \textit{et al.} \cite{ou2021active} apply the combination of network embedding and DRL to active screening problem for recurrent disease prevention on a social network under the SIS model which is similar to the IC model, where the GCNs \cite{welling2017semi} are used for generating node embedding.
        
        
\subsection{Online Influence Maximization}
    Most existing IM algorithms are applied to social networks with known graph structures and hand-crafted diffusion probabilities of edges. However, in reality, the diffusion probabilities are usually implicit and costly to acquire, which issues the demand for learning the influence probabilities. It is referred to as the online IM problem \cite{lei2015online}, which assumes that the diffusion model is known but the model parameters are unknown. Here, we review some ML-based online IM algorithms.
        
    
    \subsubsection{Edge Probability Prediction}
    Bourigault \textit{et al.} \cite{bourigault2016representation} learn the parameters of IC model by considering the similarity between node embedding vectors. They first embed the social network into a latent space to get node embedding vectors based on the community structures. Then the influence diffusion probabilities of IC model are computed by using the embeddings' distance information. Lee \textit{et al.} \cite{lee2017identifying} propose to construct brain networks by identifying edges contrasting between groups. They combine multiple CNNs in their model to learn an weighting function for each edge according to time series samples, which captures the correlations between brain regions and infers weighted undirected graphs for network construction.
    
    \subsubsection{Cascade Increment Prediction}
    Li \textit{et al.} \cite{li2017deepcas} learn an end-to-end model called \textit{DeepCas} to predict the influence cascade growth. \textit{DeepCas} first samples nodes sequences by using random walk method from cascade subgraphs at each time step. Then feed those sequences into a Gated Recurrent Unit (GRU) \cite{chung2014empirical} model with attention mechanism to learn the embeddings of the subgraphs. Based on the subgraph embeddings, an MLP is utilized for the prediction of cascade size. Liu \textit{et al.} \cite{liu2019predicting} propose a GCN-based model to predict the influence probabilities by taking into account both node and edge features as well as cascade information, which is trained with subgraphs for efficient concern and achieves $10\mbox{-}20$\% higher spread scores in the IM problem than a robust baseline methods. Chen \textit{et al.} \cite{chen2019information} propose a Recurrent Cascades Convolutional Networks (CasCN) to predict the increment size of cascade next. CasCN samples sub-cascade graphs instead of random walk sequences for preservation of cascade dynamics and local structures. GCNs and LSTMs are used by CasCN to extract topological and temporal features from the cascade subgraphs, and then produce an aggregated embedding vector which is fed into an MLP for the prediction of the cascade increment within a fixed time interval. It shows that CasCN outperforms the SOTA baselines such as \textit{node2vec}, \textit{DeepCas} and Topo-LSTM in cascade prediction. Also, Zhao \textit{et al.} \cite{zhao2022predicting} adopt similar idea of \cite{chen2019information} to build up a model named CasTCN for prediction of increment size of cascade.
    
    \subsubsection{Node Probability Prediction}
    Wang \textit{et al.} \cite{wang2017topological} claim that \textit{DeepCas} under-explores cascades by considering deficient traits of cascade structures. To overcome this, they propose an LSTM-based model named Topo-LSTM to predict the probability of a node being activated next given previous cascades. Topo-LSTM introduces diffusion topology, which is a DAG depicting cascade till the current time step, incorporating rich information of activation attempts and activation outcomes over time step series. Topo-LSTM first learns the representation of nodes in the dynamic DAGs, which then are used for predicting activation possibilities by measuring the embedding similarity. Qiu \textit{et al.} \cite{qiu2018deepinf} focus on predicting the node-level influence probabilities according to both network structures and node-specific features. They propose an end-to-end supervised model, \textit{DeepInf}, which integrates node embedding, GCN and GAT techniques to predict activation status of a node. By considering the activation status of the neighbors and the local subgraph structure of the target node, \textit{DeepInf} can output the probability of the target node being activated, and outperforms the baselines with hand-crafted features on public social network datasets. Rather than assuming known network structure like Topo-LSTM, Islam \textit{et al.} \cite{islam2018deepdiffuse} consider the problem of predicting the the activation timing and the probability of the activation of a certain node only based on observed cascades, without knowing its neighborhood. The proposed model called \textit{DeepDiffuse} is composed of two main LSTM-based deep network parts in the architecture: one generates cascade embeddings, and the other calculates the next activated node and timing, which has better performance than \textit{DeepCas} and Topo-LSTM.

\subsection{IM-based Problems}
    Recently, there emerge deep learning methods that solve optimization problems on social networks from an IM perspective. Cor{\`o} \textit{et al.} \cite{coro2019recommending} integrate the IM problem into a link recommendation, which maximizes the social influence of a given targeted users' set by suggesting a fixed number of new connections. They propose a greedy-based algorithm that can give a constant approximation ratio and can be used in large-scale real-world social networks. Zhang \textit{et al.} \cite{zhang2021grain} advocate an IM-based method to enhance the efficiency of graph-structured data selection for GNN. Yu \textit{et al.} \cite{yu2021self} propose a CNN-based framework for recommendation dealing with hypergraph which can model the high-order relations of users. Self-supervised learning can be integrated to compensate for the aggregating loss in the model by maximize mutual information within the hypergraph. Zhu \textit{et al.} \cite{zhu2021transfer} propose an approach motivated by the IM problem to facilitate the training of pre-trained transferable GNN, then produce a guidance of transferability of GNNs between source and target graphs. Zhou \textit{et al.} \cite{zhou2021overcoming} study an IM-based experience replay mechanism which selects experience nodes based on influence function in order to alleviate the forgetting problem in generic GNNs on learning a series of graph-related tasks. Ma \textit{et al.} \cite{ma2022adversarial} apply an IM-based method to help construct effective attack strategy on several GNN architectures.
    
    Thus, the IM problem can also be used as a tool to solve other ML problems or in data science field, which is very interesting and give a new research direction.

\section{Research Challenges and Directions}
    In this section, we summarize some significant research challenges in this area, including the difficulties for ML-based approaches to solve combinatorial optimization problems and the difficulties for ML-based approaches to solve the IM problem. Finally, we give several directions for the ML-based IM algorithms in the future.
    
\subsection{ML-Based Algorithms to Solve the CO Problem}
    Because the solution quality of a CO problem directly output by deep neural network cannot be compared with that of professional solvers, and the search space is too large, some auxiliary measures are needed to further improve the performance, such as beam search and local search, thus we cannot completely get rid of the dependence on traditional algorithms. It indicates that the current deep models still have a lot of room for improvement. Besides, in the training of DRL algorithms, they usually design an end-to-end model which includes graph embedding and RL model. However, commonly used training algorithms, such as REINFORCE, DQN family, and Actor-Critic, have the defects of low sampling efficiency and slow convergence. Thus, how to improve the graph embedding and RL training algorithm according to the actual needs of a specific CO problem is a very important challenge. 
    
    As we know, the IM problem is a specific CO problem on graphs. Thus, the main challenges of ML-based CO algorithms also apply to the IM problem. We summarize them from five different perspectives.
    \begin{itemize}
        \item \textbf{Scalability:} It refers to the performance when running on large datasets with the same distribution. This property is very important for the IM problem because the social network is usually large-scale, including hundreds of millions of users. However, current ML-based algorithms cannot guarantee the quality of the solution when the graph is large. Compared to small-scale networks, the quality of the solution decreased obviously.
        \item \textbf{Adaptivity:} It refers to whether the algorithm can adapt to perturbation or adjustment in the problem. Back to the IM problem, there are a lot of related variants, for example, change the objective function in a small range, increase or decrease constraints, etc. An ML-based framework can perfectly solve the vallina IM problem but cannot effectively adapt to variant IM problems.
        \item \textbf{Generalizability:} It refers to the performance when running on unseen instances, which is the reason why we attempt to use ML-based algorithms to solve the IM problem. In the IM problem, this can avoid repeating the process of seed set selection when the graph changes dynamically. Thus, we want to let the ML-based algorithms be trained by using a number of small-scale graphs, and it can work well in large-scale new graphs. Chen \textit{et al.} \cite{chen2022touplegdd} have made efforts to this end, and they improve and validate the generalization ability of DRL algorithm with multi-layer GNNs.
        \item \textbf{Efficiency:} It refers to the running time. In ML-based algorithms, its running time can be divided into training time and prediction time. The training time mainly depends on hardware conditions, dataset size, and convergence speed of the algorithm. As an algorithm designer, its main goal and difficulty lie in optimizing the convergence speed of algorithms, and the convergence speed of existing RL algorithms is also a big challenge, requiring multiple interactions with the environment. The prediction time is much shorter than the training time, but there is room for further improvement due to the large scale of social networks. It is the key to whether its algorithm can be applied to real-time environments.
        \item \textbf{Automation:} It refers to whether ML-based algorithms can automatically achieve scalability, adaptivity, and generalizability. For example, it can automatically revise parameters or structures to adapt to the change of network scale, disturbance of problems, and new unknown networks without the need for redesigning algorithms or retraining. This automatic framework not only meets all requirements of real applications, but also has intelligent adaptability to the environment. This is also the ultimate goal of artificial intelligence.
    \end{itemize}
    
    ML-based algorithms to solve CO problems has developed rapidly in recent years, but they are still in their infancy. Up to now, we can only deal with some simple and typical CO problems, whose scope of application is far less than the traditional optimization algorithm. Although the above goals have not been achieved at present: scalability, adaptivity, generalizability, efficiency, and automation, they points out a way for future research. Using this research framework to solve the IM problem can effectively alleviate the bottleneck faced by traditional algorithms, but it leads to a lot of new issues needed to be solved.
    
\subsection{ML-Based Algorithms to Solve the IM Problem}
    The IM problem is a typical subset selection problem, where the construction of the solution has nothing to do with the order of nodes. Thus, we can adopt the GNNs-based RL framework to effectively get a feasible solution. However, things are not easy because the IM problem has its unique difficulties, which leads to the inability to widely adopt the ML-based algorithms. We discuss from the following points.
    
    \begin{itemize}
        \item \textbf{Diversified Diffusion Models:} The current DRL framework to solve the IM problem adopts graph embedding to represent the states of nodes, which is shown as Eqn. (\ref{eq17}). It relies on the weigh $w(v,u)$ of each edge $(v,u)\in E$. It is perfect to be used in classic IC model and LT model because there is a weight associated with each edge. However, these simple diffusion models cannot meet the needs of real applications. The real diffusion process is usually obtained by mining diffusion datasets, thus the diffusion model formed in this way cannot be simply represented by edge weights. Therefore, it brings difficulties to the training of node representation and reinforcement learning. To address this challenge, we can consider studying how to combine deep neural networks with attention mechanism. Through the attention mechanism, we can directly train and quantify the influence of a node's neighbors on it. Moreover, some studies have explored data-driven non-modeling methods for optimization problems \cite{tong2021usco} \cite{tong2022social} \cite{wang2022learnability}, by abandoning the objective function and training a solver from data of input-solution pairs. We can borrow ideas from these methods to solve the IM problem without assuming the latent diffusion model.
        \item \textbf{Training Efficiency:} The training process of the current DRL framework is shown in Eqn. (\ref{eq17}) and Eqn. (\ref{eq18}). A serious defect here is that the embedding $\mu^{i+1}_v$ depends on $a_v$, whether or not the node $v$ is in current seed set $S$. Thus, when we update the seed set $S$ in each iteration, we have to embed all nodes in the graph again. This is very time consuming and inefficient. When the network scale is large, the training time will become unacceptable. Thus, this scheme is not perfect. A feasible idea is to study how to encode a set, that is, the seed set in the IM problem. Here, we can consider separating the partial solution $S$ from the encoding of candidate nodes $v$, that is, the embedding of a node $v$ does not depend on whether it is in the partial solution $S$, thereby reducing the update times of graph embedding and the time complexity of training. Then, it can make $Q$ value satisfy
        \begin{equation}
            Q(enc(S),enc(v);\Theta)\propto\sigma_\Omega(v|S).
        \end{equation}
        \item \textbf{Reward Estimation:} In the process of designing RL algorithm, a key step is to define the reward function. For the IM problem, when adopt action $v$ based on current state $S$, the reward function is usually defined as the marginal gain, $R(S,v)=\sigma_\Omega(v|S)$. However, we have known that this value is very difficult to accurately compute. Now, this value can be estimated by using Monte Carlo simulation \cite{wang2021reinforcement} \cite{li2022piano}, which greatly increases the training time. Here, we can adopt some techniques used by traditional algorithms, such as proxy-based and sampling-based approaches to accelerate this estimation process. The \#P-hardness of computing the influence spread in the IM problem is also a huge challenge in ML-based algorithms, which leads to its inability to completely get rid of its dependence on traditional algorithms, and makes the scalability and generalization ability worse. Thus, to study how to overcome the \#P-hardness of computing the influence spread in ML-based algorithm is a feasible research direction.
        \item \textbf{Variant Influence Maximization:} Shown as Section \uppercase\expandafter{\romannumeral7}, there are a lot of variants based on the IM problem, which are mainly reflected by the following differences. Generally, they have different objective functions designed according to various application scenarios, which usually cannot guarantee to meet some good properties, such as monotonicity and submodularity. Perhaps, there have more complex constraints or a higher requirement for real-time and generalization ability. For the variant IM problems, they usually have complex characteristics such as multi-objective, multi-constraints, and non-static, which cannot be effectively approximated by traditional algorithms. Thus, we can consider using ML-based algorithms to study these kinds of problems with bad properties, and make full use of the advantages of ML-based algorithms to solve the problems that traditional algorithms are difficult to solve.
    \end{itemize}
    
    According to the above analysis, we can see that there are still many challenges in solving the IM problem in large-scale social networks by using existing ML-based approaches. Up to now, we can only utilize a simple DRL framework to deal with the vanilla IM problem under the IC model in medium-sized social networks, and has great limitations. It is mainly reflected in the fact that it cannot adapt to diversified diffusion models in real applications, the training efficiency based on GNNs is not acceptable, the calculation of reward functions cannot overcome the \#P-hardness of computing the influence spread, and cannot adjust to more complicated variant problems. Because of this, it also brings new research opportunities and vitality to this field. To deal with these challenges, we need to make efforts from the basic theory of machine learning and the modeling of social network applications, and combine them effectively.

\subsection{Summary and Future Directions}
    To deal with the IM problem and its variants by using ML-based methods, we need to make efforts in two aspects. One is the basic research of solving combinatorial optimization problems by machine learning, then it tries to design new frameworks or improve the existing ML-based frameworks to enhance scalability, adaptivity, generalizability, efficiency, and automation. The GNN-based RL will be the mainstream framework in the future. Compared with supervised learning, it does not need a lot of datasets. It is very difficult to produce a large number of datasets in CO problems. In addition to relying on traditional RL, one of the key factors affecting algorithm performance when dealing with graph problems is graph representation, which largely determines scalability and generalizability. Here, we should try to combine with the latest deep learning technologies, such as attention mechanism and transformer. Finally, we should emphasize that adopting ML-based methods does not mean completely giving up traditional methods, where some ideas from approximation or heuristic algorithms are still worth learning. As an instance, using some simple heuristic rules to prune the search space will make the subsequent operation more efficient. How to fully combine the two kinds of strategies and complement each other's advantages is also an important development direction in the future.
    
    The other is how to apply the existing ML-based CO algorithms to solve the IM problem and its variants. The IM problem in social networks has its unique characteristics, which leads to the general solution framework for CO problems based on machine learning not fully adapting to it. It is mainly embodied in the following three points. First, social networks are very large. This is very demanding of scalability and efficiency. The quality of the solutions output by the current ML-based CO algorithms on large-scale networks is poor, and the training time of the scheme relying on the graph neural network model is too long, which leads to its applications limiting the network scale. Second, social applications at present require high generalization ability of algorithms. This is due to the dynamic change of network topology caused by the number of users and the dynamic change of friendships among users. Third, computing the influence spread is not the valuable oracle. This is because computing the spread is \#P-hard under classic diffusion models, which brings great difficulties to the training of ML-based algorithms, and it cannot completely get rid of dependence on traditional search algorithms. Fourth, variant IM problems totally inherit the above characteristics, and the objective functions and constraints are more complex. This is very demanding of adaptivity and automation.
    
    Therefore, we can seek plenty of innovation and research opportunities based on the above two major aspects. Here, we can not only study the basic theory of solving CO problems by adopting ML-based approaches, but also study the transfer of existing theories to the IM problem and its variants. Our goal is to construct an ML-based solution framework for the IM problem to meet the practical application requirements, including scalability, adaptivity, generalizability, efficiency, and automation. Finally, they jointly promote the progress of theory and technology in this field.

        
        

\section{Conclusion}
    In this paper, we present a comprehensive survey on the IM problem from an ML-based combinatorial optimization perspective. In order to compare ML-based algorithms with traditional approaches, we first summarize three kinds of traditional approaches: simulation-based, proxy-based, and sampling-based approaches, and point out their respective defects. Then, we survey two kinds of commonly used framework that can be used to solve combinatorial optimization problems: Ptr-Nets Based RL framework and GNNs Based RL Framework. We discuss their basic structures and characteristics in detail, and focus on summarizing the principles and technologies related to the IM problem. This paves the way for solving the IM problem by using ML-based methods. Furthermore, we cover almost all the ML-based IM algorithms up to now, focusing on several pioneering DRL solution frameworks, and summarizing the variant IM problems and their corresponding ML-based algorithms. Here, we elaborately explain their working principle and how to break through the bottleneck of traditional algorithms. Finally, we summarize the existing difficulties and future research directions in this field from two aspects.
    
    In summary, machine learning, as a representative of emerging technology, is a hot topic at present. Its application in solving algorithm problems in social networks has broad application prospects and commercial value, and it is expected to solve the dilemma faced by traditional algorithms. Our survey will give relevant researchers a new perspective to look at the IM problem, and can be used as a learning material to understand the latest ML-based development and research trends in the field of social network and algorithm design, thus laying a good start for the research in this field.

\begin{acks}
This work was supported in part by the National Natural Science Foundation of China (NSFC) under Grant No. 62202055, the Start-up Fund from Beijing Normal University under Grant No. 310432104, the Start-up Fund from BNU-HKBU United International College under Grant No. UICR0700018-22, and the Project of Young Innovative Talents of Guangdong Education Department under Grant No. 2022KQNCX102.
\end{acks}

\bibliographystyle{ACM-Reference-Format}
\bibliography{references}

\end{document}